\documentclass[
  aps,
  prc,
  reprint,
  groupedaddress,
  amsmath,
  amssymb,
  noshowpacs,
  floatfix,
  nofootinbib
]{revtex4-1}

\usepackage[english]{babel}
\usepackage{aas_macros} 
\usepackage{xspace}
\usepackage{color}
\usepackage{graphicx}
\usepackage{bm}
\usepackage{dcolumn}
\usepackage[colorlinks=true,allcolors=blue]{hyperref}
\usepackage{microtype}

\newcommand{\uncaddr} {
  Department of Physics and Astronomy, CB 3255,
  University of North Carolina, Chapel Hill, NC 27599-3255}
\newcommand{\ncsuaddr}{
  Department of Physics, NC State University, Box 8202,
  Raleigh, NC 27695}

\newcommand*{\asimwk}{\ensuremath{A \simeq 80}\xspace}
\newcommand*{\asimre}{\ensuremath{A \simeq 160}\xspace}
\newcommand*{\moetal}{M\"oller \emph{et al}} % E.g. \moetal.\

\newcommand*{\cs}{\ensuremath{C_{{1}{0}}^{s}}\xspace}
\newcommand*{\skop}{{SkO$^\prime$}\xspace}
\newcommand*{\ndsk}{SkO$'$-Nd\xspace}
\newcommand*{\uneo}{\textsc{unedf1-hfb}\xspace}
\newcommand*{\pounders}{\textsc{pound}er\textsc{s}\xspace}

\newcommand*{\calH}{\ensuremath{\mathcal{H}}}
\newcommand*{\calR}{\ensuremath{\mathcal{R}}}
\newcommand*{\bbW}{\ensuremath{\mathbb{W}}}
\newcommand*{\bbR}{\ensuremath{\mathbb{R}}}
\newcommand*{\bbH}{\ensuremath{\mathbb{H}}}
\newcommand*{\bbF}{\ensuremath{\mathbb{F}}}

\newcommand*{\bbr}{\ensuremath{\mathbb{R}}}
\newcommand*{\bbh}{\ensuremath{\mathbb{H}}}
\newcommand*{\bbf}{\ensuremath{\mathbb{F}}}

\newcommand*{\efa}{\ensuremath{\mathrm{EFA}}}
\newcommand*{\dop}{\ensuremath{\hat{\mathcal{D}}}}
\newcommand*{\comm}[2]{\ensuremath{\big[#1 ,\,#2\big]}}

\newcommand*{\mel}[3]{\ensuremath{\langle #1 \,|\, #2 \,|\, #3 \rangle}}
\newcommand*{\rmel}[3]{\ensuremath{\langle #1 \,||\, #2 \,||\, #3 \rangle}}
\newcommand*{\abs}[1]{\ensuremath{\left|#1\right|}}
\newcommand*{\emiwt}{\ensuremath{e^{-i\omega t}}}
\newcommand*{\epiwt}{\ensuremath{e^{i\omega t}}}
\newcommand*{\delmnh}{\ensuremath{\Delta M_{n-H}}\xspace}
\newcommand*{\cg}[6]{\ensuremath{(\,{#1}\,{#2}\;{#3}\,{#4}\,|\,{#5}\,{#6}\,)}}
\newcommand*{\ket}[1]{\ensuremath{|\,#1\,\rangle}}
\newcommand*{\bra}[1]{\ensuremath{\langle\,#1\,|}}
\newcommand*{\epn}{\ensuremath{E_0(\mathrm{pn})}}
\newcommand*{\dav}{\ensuremath{\tilde{\Delta}^{(3)}}}
\newcommand*{\ddd}{\ensuremath{\Delta^{(3)}}}

\newcommand*{\ex}{\ensuremath{E_\mathrm{ex}}}
\newcommand*{\ccc}{\ensuremath{c^{\dagger}}}
\newcommand*{\cca}{\ensuremath{c^{\vphantom{\dagger}}}}
\newcommand*{\aac}{\ensuremath{\alpha^{\dagger}}}
\newcommand*{\aaa}{\ensuremath{\alpha^{\vphantom{\dagger}}}}
\newcommand*{\hfb}{\ensuremath{\Phi}}
\newcommand*{\khfb}{\ensuremath{\ket{\hfb}}}
\newcommand*{\bhfb}{\ensuremath{\bra{\hfb}}}
\newcommand*{\ev}[1]{\ensuremath{\langle #1 \rangle}}

\DeclareMathOperator{\Tr}{Tr}
\DeclareMathOperator{\tr}{tr}
\begin{document}

\title{\boldmath
  Beta decay of deformed \textit{r}-process nuclei near $A\!=\! 80$ and $A\!=\! 160$,\\
  including odd-$A$ and odd-odd nuclei, with the Skyrme finite-amplitude method}

\author{T.~Shafer}
\email{tom@tshafer.com}
\affiliation{\uncaddr}

\author{J.~Engel}
\email{engelj@physics.unc.edu}
\affiliation{\uncaddr}

\author{C.~Fr\"{o}hlich}
\affiliation{\ncsuaddr}

\author{G.~C.~McLaughlin}
\affiliation{\ncsuaddr}

\author{M.~Mumpower}
\affiliation{Department of Physics,
  University of Notre Dame, Notre Dame, IN 46556}
\affiliation{Theory Division,
  Los Alamos National Laboratory, Los Alamos, NM 87544}

\author{R.~Surman}
\affiliation{Department of Physics,
  University of Notre Dame, Notre Dame, IN 46556}

\date{\today}

\begin{abstract}
After identifying the nuclei in the \asimwk and \asimre regions for which
$\beta$-decay rates have the greatest effect on weak and main \textit{r}-process
abundance patterns, we apply the finite-amplitude method (FAM) with Skyrme
energy-density functionals (EDFs) to calculate $\beta$-decay half-lives of those
nuclei in the quasiparticle random-phase approximation (QRPA).  We use the equal
filling approximation to extend our implementation of the charge-changing FAM,
which incorporates pairing correlations and allows axially symmetric
deformation, to odd-$A$ and odd-odd nuclei.  Within this framework we find
differences of up to a factor of seven between our calculated $\beta$-decay
half-lives and those of previous efforts.  Repeated calculations with \asimre
nuclei and multiple EDFs show a spread of two to four in $\beta$-decay
half-lives, with differences in calculated $Q$ values playing an important role.
We investigate the implications of these results for \textit{r}-process
simulations. 
%The extension of the FAM to odd-$A$ and odd-odd nuclei will enable full Skyrme
%%QRPA $\beta$-decay calculations across the nuclear chart.
\end{abstract}


\maketitle

% ------------------------------------------------------------------------------
% ------------------------------------------------------------------------------

\section{\label{sec:introduction}Introduction}

% r-process basics
The solar abundances of nuclei heavier than iron, on the neutron-rich side of
stability, have traditionally been attributed to rapid neutron-capture, or
\textit{r}-process, nucleosynthesis \cite{BBFH}.  The three largest abundance
peaks in the solar pattern, at $A\sim 80$, 130, and 195, are associated with the
closed neutron shells at $N=50$, 82, and 126, suggesting that astrophysical
conditions of increasing neutron-richness are responsible for each.  A
smaller fourth abundance peak in the rare-earth elements ($A\sim 160$)
is also formed in neutron-rich environments.  Observational data from meteorites
and metal-poor halo stars confirm the separate origins for $70\lesssim A
\lesssim 120$ (``weak'') and $A>120$ (``main'') \textit{r}-process nuclei and
provide hints of the nature of the \textit{r}-process astrophysical site, though
the exact site (or sites) has not yet been definitively pinned
down \cite{Arn07}.

% Importance of nuclear data for the r process
In principle the \textit{r}-process sites can be identified by comparing
simulations of prospective astrophysical environments with observational data
from the solar system and other stars (see, e.g., Ref.\ \cite{Shibagaki+16}).
The precision of \textit{r}-process abundance predictions, however, is limited
by our incomplete knowledge of properties---such as masses, reaction rates, and
decay lifetimes---of nuclei on the neutron-rich side of
stability \cite{mumpower16}.  It is particularly important that we better
determine decay lifetimes, since \textit{r}-process nuclei are built up via a
sequence of captures and $\beta$ decays.  Thus, $\beta$-decay lifetimes
determine the relative abundances of the nuclei along the \textit{r}-process
path \cite{BBFH,Arn07,Mum14} and the overall timescale for neutron
capture \cite{moller97,engel99}.  At late times, as nuclei move back from the
\textit{r}-process path towards stability and the last remaining neutrons are
captured, the lifetimes determine the shape of the final abundance pattern
\cite{Mum14,Caballero+14}. Finally, for a weak \textit{r} process $\beta$-decay
rates control the amount of material that remains trapped in the $A\sim80$ peak
and the amount that moves to higher mass numbers \cite{Mad12}, i.e.\ they
determine where the weak \textit{r} process terminates. For all these reasons,
an accurate picture of the $\beta$ decay of neutron-rich nuclei is crucial for
the accuracy of \textit{r}-process simulations.

% Importance of theoretical calculations + our approach
Although many $\beta$-decay lifetimes have been measured (see, e.g., Refs.\
\cite{Hos05,Hos10,Mad12,Maz13,Mie13}), most of the nuclei populated during the
\textit{r} process remain out of reach. Simulations must therefore rely on calculated
lifetimes.  The most widely-used sets of theoretical rates are from gross
theory \cite{Takahashi+69,Koyama+70,Takahashi71} and from an application of the
quasiparticle random-phase approximation (QRPA) within a macroscopic-microscopic
framework \cite{moller97,moller03} that employs gross theory for first-forbidden
transitions.  Here we use a fully microscopic Skyrme QRPA, implemented through
the proton-neutron finite-amplitude method (pnFAM) \cite{Mustonen14} and
extended to treat odd-$A$ and odd-odd nuclei (hereafter ``odd'' nuclei) in the
equal-filling approximation (EFA) \cite{Perez-Martin08}.  We can now use
arbitrary Skyrme energy-density functionals (EDFs) to self-consistently compute
$\beta$-decay rates of both even-even and odd axially-symmetric nuclei,
including contributions of both allowed ($J^\pi=1^+$) and first-forbidden
($J^\pi=0^-$, $1^-$, or $2^-$) transitions.

% Choice of nuclei
We evaluate lifetimes for key \textit{r}-process nuclei in two highly populated
regions of the abundance pattern: the large maximum at $A\sim 80$ and the
smaller rare earth peak at $A\sim 160$. In a main \textit{r}-process, the rare
earth peak forms in a different way than do the large peaks at $A\sim 130$ and
195, both of which originate from long-lived ``waiting points'' near closed
neutron shells at $N=82$ and 126. The rare earth peak, by contrast, forms during
the late stages of the \textit{r} process, as $\beta$ decay, neutron capture,
and photo-dissociation all compete with one another and the \textit{r}-process
path moves toward stability \cite{surman97,Mumpower11}.  The $A\sim 160$
abundance peak is thus useful for studying the main \textit{r}-process
environment \cite{Mumpower12,Mumpower+16b}. The $A\sim 80$ region is not so
clearly related to the main \textit{r} process.  In fact, nuclei with
$70\lesssim A\lesssim 120$ can be created in a variety of nucleosynthetic
processes, ranging from the neutron-rich weak \textit{r} process to the
proton-rich $\nu p$ process \cite{Frohlich06,Pruet06,Wanajo06} (see also Refs.\
\cite{Thi11,Arc11}).  Untangling the various contributions to these elements
requires rigorous abundance pattern predictions, which in turn require a better
knowledge of still unmeasured $\beta$-decay half-lives. 

% What we do here
In this paper, we aim to study and improve \textit{r}-process abundance
predictions for both weak \textit{r}-process nuclei and the rare-earth elements
by identifying and recalculating key $\beta$-decay rates. We begin in Sec.\
\ref{sec:ns} by reviewing the pnFAM and then discussing our extension to odd
nuclei.  In Sec.\ \ref{sec:hl}, guided by \textit{r}-process sensitivity
studies, we calculate $\beta$-decay rates separately for the important isotopes
in the two mass regions (after optimizing the Skyrme EDF separately for each
region).  We also examine the effect on $\beta$-decay half-lives of varying the
Skyrme EDF in rare-earth nuclei.  Finally, in Sec.\ \ref{sec:rp} we discuss the
impact of our $\beta$-decay rates on \textit{r}-process abundances.
Sec.\ \ref{sec:conclusion} is a conclusion.

% ------------------------------------------------------------------------------
% ------------------------------------------------------------------------------

\section{\label{sec:ns}Nuclear structure}

% ------------------------------------------------------------------------------

\subsection{\label{sec:ns-fam}The proton-neutron finite-amplitude method}

% Review of the FAM literature and our extension
The finite-amplitude method (FAM) is an efficient way to calculate strength
distributions in the random-phase approximation (RPA) or the QRPA\@.
Nakatsukasa \textit{et al}.\ first introduced the FAM to calculate the RPA
response of deformed nuclei \cite{Nakatsukasa07} with Skyrme EDFs, and the
method has been rapidly extended to include pairing correlations in Skyrme QRPA,
both for spherical \cite{Avogadro11} and axially-deformed
nuclei \cite{Kortelainen15}, and to include similar correlations in the
relativistic QRPA \cite{Niksic13}.  Ref.\ \cite{Mustonen14} applied the same
ideas to charge-changing transitions, in particular those involved in $\beta$
decay; the resulting method is called the pnFAM\@.  Like the FAM implemented in
Ref.\ \cite{Kortelainen15}, the pnFAM computes strength functions for
transitions that change the $K$ quantum number by arbitrary (integer) amounts in
spherical or deformed superfluid nuclei.  

The first work with the pnFAM focused on the impact of tensor terms in Skyrme
EDFs \cite{Mustonen14}.  More recently, the authors of Ref.\ \cite{Mustonen16}
used the method to constrain the time-odd part of the Skyrme EDF and compute a
$\beta$-decay table that includes the half-lives of 1387 even-even nuclei.  We
leave most details of the pnFAM itself to these references, but repeat the main
points here in anticipation of the extension to odd nuclei in
Sec.\ \ref{sec:ns-efa}.

% High-level HFB definitions
QRPA strength functions are related to the linear time-dependent response of the
Hartree-Fock-Bogoliubov (HFB) mean field (see, e.g.,
Refs.\ \cite{Ring05,Blaizot85} for a discussion). The static HFB equation can be
written as 
\begin{equation} \label{eq:hfb-comm-sp}
\comm{\calH_0}{\calR_0} = 0,
\end{equation}
where (e.g., for protons or neutrons)
\begin{equation} \label{eq:hfb-rh-sp}
\calR_0 = \begin{pmatrix} \rho_0 & \kappa_0 \\ -\kappa_0^* & 1-\rho_0^* \end{pmatrix}, \quad
\calH_0 = \begin{pmatrix} h_0 & \Delta_0 \\ -\Delta_0^* & -h_0^* \end{pmatrix}.
\end{equation}
In Eq.\ \eqref{eq:hfb-rh-sp}, $\calR_0$ is the generalized static density (the
subscript 0 indicates a static quantity), built from the single-particle density
$\rho_0$ and the pairing tensor $\kappa_0$ (see, e.g., Ref.\ \cite{Ring05}), and
$\calH_0$ is the static generalized mean field, built from the static mean field
$h_0$ and the static pairing field $\Delta_0$.  The generalized mean field
$\calH_0$ depends on both $\rho_0$ and $\kappa_0$ and is usually written
$\calH_0[\calR_0]$.  The matrices $\calR_0$ and $\calH_0$ are diagonalized by a
unitary Bogoliubov transformation,
\begin{equation}
\label{eq:W-def}
\bbW = \begin{pmatrix} U & V^* \\ V & U^* \end{pmatrix},
\end{equation}
which connects the set of single-particle states (created by $\ccc_k$) in which
the problem is formulated to a set of quasiparticle states (created by
$\aac_\mu$):
\begin{equation}
\begin{pmatrix} \cca \\ \ccc \end{pmatrix} =
\begin{pmatrix} U & V^* \\ V & U^* \end{pmatrix}
\begin{pmatrix} \aaa \\ \aac \end{pmatrix}.
\end{equation}
Thus, the transformed generalized density and mean field,
\begin{equation}
\bbr_0 \equiv \bbW^\dagger \calR_0 \bbW \,, \quad
\bbH_0 \equiv \bbW^\dagger \calH_0 \bbW \,,
\end{equation}
are in the quasiparticle basis and have the diagonal form 
\begin{equation} \label{eq:rh-qp-def}
\bbR_0 = \begin{pmatrix} 0 & 0 \\ 0 & 1 \end{pmatrix}, \quad
\bbH_0 = \begin{pmatrix} E & 0 \\ 0 & -E \end{pmatrix}.
\end{equation}

In the pnFAM we solve the small-amplitude time-dependent HFB (\mbox{TDHFB}) equation,
\begin{equation} \label{eq:tdhfb}
i \dot\bbr(t) = \comm{\bbh[\bbr(t)] + \bbf(t)}{\bbr(t)},
\end{equation}
where $\bbf(t)$ is a time-dependent external field that changes neutrons into
protons or vice versa.  Equation \eqref{eq:tdhfb} determines the oscillation of
the generalized density around the static solution $\bbR_0$ of Eq.\
\eqref{eq:hfb-comm-sp}; for external fields proportional to a small parameter
$\eta$, a first-order expansion $\bbr(t) \approx \bbr_0 + \eta \delta\bbr(t)$ is
sufficient to describe the behavior of the nucleus.  It leads to the
linear-response equation:
\begin{equation} \label{eq:hfblr}
i\delta\dot\bbr(t) = \comm{\bbh_0}{\delta\bbr(t)}
  + \comm{\delta\bbh(t) + \bbf(t)}{\bbr_0} \,.
\end{equation}
Here $\delta\bbh(t)$ and $\delta\bbr(t)$ are the first-order changes in the
generalized mean field and density.  

If the perturbing field oscillates at a frequency $\omega$, the resulting
generalized density can be written in the form: 
\begin{equation} \label{eq:delta-r-ansatz}
\delta\bbr(t) = \delta\bbr(\omega) \emiwt + \delta\bbr^\dagger(\omega) \epiwt
\,,
\end{equation}
with
\begin{equation} \label{eq:pnfam-even-dr-omega}
\delta\bbr(\omega)
\equiv \begin{pmatrix} 0 & X(\omega) \\ -Y(\omega) & 0 \end{pmatrix},
\end{equation}
where the requirement that $\bbr(t)$ remain projective ($\bbr^2 = \bbr$) forces
the diagonal blocks to be zero \cite{Niksic13}.  The time-dependent generalized
Hamiltonian also oscillates harmonically, with 
\begin{equation}
\delta\bbH(\omega)
= \begin{pmatrix}
  \delta H^{11}(\omega) & \delta H^{20}(\omega) \\
  -\delta H^{02}(\omega) & -\delta H^{\overline{11}}(\omega) \,.
\end{pmatrix},
\end{equation}
The superscripts on the blocks in $\delta\bbH$ refer to the number of
quasiparticles created and destroyed by the corresponding block Hamiltonian.

Putting everything together in Eq.\ \eqref{eq:hfblr} (including the oscillating
external field $\bbF(t)$, which we have not written out explicitly here), and
evaluating the commutators, one obtains the pnFAM equations \cite{Mustonen14}:
\begin{subequations}\label{eqs:pnfam-eqs}
\begin{alignat}{2}
X_{\pi\nu}(E_\pi + E_\nu - \omega) &+ \delta H^{20}_{\pi\nu}(\omega) &=& -F^{20}_{\pi\nu},\\
Y_{\pi\nu}(E_\pi + E_\nu + \omega) &+ \delta H^{02}_{\pi\nu}(\omega) &=& -F^{02}_{\pi\nu},
\end{alignat}
\end{subequations}
where $\pi$ and $\nu$ label proton and neutron states, and $E_\pi$ and $E_\nu$
are single-quasiparticle energies.  Eqs.\ \eqref{eqs:pnfam-eqs} can be put into
matrix-QRPA form \cite{Avogadro11}, but they are more easily solved directly
(through iteration) \cite{Nakatsukasa07,Avogadro11,Mustonen14}.  The FAM
transition strength is then just given by $S(F;\omega) = \tr \bbf^\dagger
\delta\bbr^{(pn)}(\omega)$ \cite{Nakatsukasa07,Avogadro11,Mustonen14}.

% ------------------------------------------------------------------------------

\subsection{\label{sec:ns-efa}The equal-filling approximation and the linear response of odd nuclei}

% Overview of the EFA
Our pnFAM code and \textsc{hfbtho}, the HFB code on which it is based, require
time-reversal-symmetric nuclear states \cite{Stoitsov13}.  To apply the FAM to
odd nuclei, the ground states of which break time-reversal symmetry, we use the
EFA\@, an ``phenomenological'' approximation, in the words of Ref.\
\cite{Perez-Martin08} in which the interaction between the odd nucleon and the
core are captured at least partially without breaking time-reversal symmetry.

In odd-nucleus density-functional theory, the ground state is typically
represented in leading order by a one-quasiparticle excitation of an even-even
core, $\ket{\Phi_\Lambda} = \aac_\Lambda\khfb$.  This state, however, produces
the time-reversal-breaking single-particle and pairing densities
\cite{Perez-Martin08,Schunck10}
\begin{subequations} \label{eq:densities-odd}
\begin{align}
\rho_{kk'}   &= (V^* V^T)_{kk'} + U_{k\Lambda} U^*_{k'\Lambda} - V^*_{k\Lambda}
V_{k'\Lambda} \,, \\
\kappa_{kk'} &= (V^* U^T)_{kk'} + U_{k\Lambda} V^*_{k'\Lambda} - V^*_{k\Lambda}
U_{k'\Lambda} \,.
\end{align}
\end{subequations}
The EFA replaces the densities in \eqref{eq:densities-odd} with new ones that
average contributions from the state $\aac_\Lambda\khfb$ and its time-reversed
partner $\aac_{\bar\Lambda}\khfb$:
% Use split: http://tex.stackexchange.com/a/119174
\begin{subequations} \label{eq:densities-efa}
\begin{align}
\begin{split} \label{eq:rho-efa}
\rho^\efa_{kk'} &= (V^* V^T)_{kk'} + \frac{1}{2} \Big(
    U_{k\Lambda} U^*_{k'\Lambda} + U_{k\bar\Lambda} U^*_{k'\bar\Lambda} \\
    & \qquad\qquad
  - V^*_{k\Lambda} V_{k'\Lambda} - V^*_{k\bar\Lambda} V_{k'\bar\Lambda} \Big)
  \,,
\end{split} \\
\begin{split}
\kappa^\efa_{kk'} &= (V^* U^T)_{kk'} + \frac{1}{2} \Big(
    U_{k\Lambda} V^*_{k'\Lambda} + U_{k\bar\Lambda} V^*_{k'\bar\Lambda} \\
    & \qquad\qquad
  - V^*_{k\Lambda} U_{k'\Lambda} - V^*_{k\bar\Lambda} U_{k'\bar\Lambda} \Big)
  \,,
\end{split}
\end{align}
\end{subequations}
where we have assumed that the state of the even-even core $\khfb$ is
time-reversal even.  The odd-$A$ HFB calculation then proceeds as usual with
$\rho \to \rho^\efa$ and $\kappa \to \kappa^\efa$ \cite{Perez-Martin08}.

% EFA justification
The EFA appears to be an excellent approximation to the full HFB solution for
odd-$A$ nuclei. Ref.\ \cite{Schunck10} contains calculations of odd-proton
excitation energies in rare-earth nuclei, in both the EFA and the blocking
approximation.  The EFA reproduces the full one-quasiparticle energies to within
a few hundred keV.  The approximation was given a theoretical foundation in
Ref.\ \cite{Perez-Martin08}, which showed that $\rho^\efa$ and $\kappa^\efa$ can
be obtained rigorously by abandoning the usual product form of the HFB solution
and instead describing the nucleus as a mixed state.  From this point of view,
the nucleus is not represented by a single state vector $\ket{\hfb_\Lambda}$ but
rather by a statistical ensemble with a density operator $\dop \equiv \exp
\hat{K}$ \cite{Perez-Martin07}:
\begin{equation} \label{eq:dop-expanded}
\begin{aligned}
\dop &= \khfb\bhfb + \sum_\mu \aac_\mu \khfb p_\mu \bhfb \aaa_\mu \\  &\qquad
  + \frac{1}{2!} \sum_{\mu\nu} \aac_{\mu}\aac_\nu \khfb p_\mu p_\nu \bhfb \aaa_\nu \aaa_\mu
  + \dots.
\end{aligned}
\end{equation}
In Eq.\ \eqref{eq:dop-expanded}, $p_\mu$ is the probability that the excitation
$\aac_\mu\khfb$ is contained in the ensemble.  Expectation values are traces
with $\dop$ in Fock space (we use `$\Tr$' for these traces and `$\tr$' for the
usual trace of a matrix),
\begin{equation}
\ev{A} = \Tr[\dop \hat A] / \Tr[\dop],
\end{equation}
so that, e.g., the particle density is
\begin{equation} \label{eq:rho-tr}
\rho_{kk'} = \Tr[\dop \ccc_{k'}\cca_k]/\Tr[\dop].
\end{equation}

An ensemble like the above is familiar from finite-temperature HFB
\cite{Goodman81}, where the quasiparticle occupations are statistical and
determined during the HFB minimization.  Ref.\ \cite{Perez-Martin08} shows that
the EFA emerges from a specific non-thermal choice of the ensemble
probabilities:
\begin{equation} \label{eq:p-mu}
p_\mu = \begin{cases}
  1, \quad & \mu \in [\Lambda, \bar\Lambda] \\
  0, & \text{otherwise} \,.
\end{cases}
\end{equation}
With these values of $p_\mu$, one finds that for an arbitrary one-body operator
$\hat{O}$,
\begin{equation} \label{eq:efa-avg-ev-eo}
\ev{\hat O}_\text{o-e} = \frac{1}{2} ( \mel{\hfb}{\aaa_\Lambda\hat O\aac_\Lambda}{\hfb}
  + \mel{\hfb}{\aaa_{\bar\Lambda}\hat O\aac_{\bar\Lambda}}{\hfb} ),
\end{equation}
and the trace in Eq.\ \eqref{eq:rho-tr} produces $\rho^\efa$ \eqref{eq:rho-efa}.
The formalism may also be applied in a straightforward way to odd-odd nuclei
as well by constructing an ensemble from the proton ($\pi$) and
neutron ($\nu$) orbitals $\Lambda_\pi$, $\bar\Lambda_\pi$, $\Lambda_\nu$, and
$\bar\Lambda_\nu$.  Then one finds that 
\begin{equation} \label{eq:efa-avg-ev-oo}
\begin{aligned}
\ev{\hat O}_\text{o-o} &= \frac{1}{2} (
  \mel{\hfb}
    {\aaa_{\Lambda_\nu} \aaa_{\Lambda_\pi} \hat O \aac_{\Lambda_\pi} \aac_{\Lambda_\nu}}
    {\hfb} \\
&\qquad+
  \mel{\hfb}
    {\aaa_{\bar\Lambda_\nu} \aaa_{\bar\Lambda_\pi} \hat O \aac_{\bar\Lambda_\pi} \aac_{\bar\Lambda_\nu}}
    {\hfb} ) \,.
\end{aligned}
\end{equation}

% Technical extension---introduction
The statistical interpretation of the EFA allows us to extend the pnFAM, which
is an approximate time-dependent HFB, to odd nuclei.  The thermal QRPA,
described in Refs.\ \cite{Sommermann83,Vautherin84,Ring84,Alasia89,Egido93}
generalizes Eq.\ \eqref{eq:hfblr} to a statistical density operator $\dop = \exp
\hat{K}$ and a thermal ensemble; here we do the same with the non-thermal
ensemble in Eq.\ \eqref{eq:p-mu}.  

% Generalized density
Of the matrices that enter the TDHFB equations \eqref{eq:tdhfb}, only the
generalized density $\bbr$ is fundamentally altered in the EFA\@; the external
field is unaffected and the ground-state Hamiltonian matrix $\bbh_0$ assumes its
usual form \cite{Perez-Martin08}.  But the replacement $\mel{\hfb}{\hat
A}{\hfb}$ by $\Tr[\dop \hat A]/\Tr[\dop]$ has implications for both the static
density $\bbr_0$ and the time-dependent perturbation $\delta\bbr(t)$.  In the
usual HFB, the definition of the generalized density in the quasiparticle basis
\cite{Ring05,Blaizot85},
\begin{equation}
\bbr = \begin{pmatrix} \ev{\aac\aaa} & \ev{\aaa\aaa} \\ \ev{\aac\aac} &
\ev{\aaa\aac} \end{pmatrix} \,,
\end{equation}
leads to the form of $\bbR_0$ in Eq.\ \eqref{eq:rh-qp-def}.  In the EFA
ensemble, however, the expectation values $\ev{\aac\aaa}$ and $\ev{\aaa\aac}$
are \cite{Perez-Martin08}:
\begin{subequations}
\begin{gather}
\ev{\aac_\nu\aaa_\mu} = \delta_{\mu\nu} f_\mu \,, \\
\ev{\aaa_\nu\aac_\mu} = \delta_{\mu\nu} (1-f_\mu) \,,
\end{gather}
\end{subequations}
leading to an $\bbr_0$ with the more general form
\begin{equation}
\bbr^\efa_0 =
  \begin{pmatrix} f & 0 \\ 0 & {1-f} \end{pmatrix} \,.
\end{equation}
The matrix $f$ is diagonal, with factors $f_\mu$ related to the $p_\mu$ \eqref{eq:p-mu} and taking on the values
\begin{equation}
f_\mu =
\begin{cases}
  \frac{1}{2}, \quad & \mu \in [\Lambda, \bar\Lambda], \\
  0, & \text{otherwise}.
\end{cases}
\end{equation}

The use of an ensemble also changes the way we calculate the response
$\delta\bbr(t)$. Following Ref.\ \cite{Sommermann83}, we consider the evolution
of the density operator under a unitary transformation $U(t) = \exp[i\eta\hat
S(t)]$.  The operator $\hat S(t)$ is undetermined, but Hermitian.  To first
order in $\eta$, the ensemble evolves as
\begin{equation}
\begin{aligned}
\dop(t) &\simeq [1 + i\eta\hat S(t)] \dop(0) [1 - i\eta\hat S(t)] \\
  &   \equiv \dop(0) + \eta \delta\dop(t),
\end{aligned}
\end{equation}
with $\delta\dop(t) = -i [\dop,\,\hat S(t)]$.  The cyclic invariance of the
trace \cite{Perez-Martin08} guarantees that $\Tr[\delta\dop(t)] = 0$.  The time
evolution of $\dop(t)$ determines the evolution of $\delta\bbR(t)$, e.g., via
\begin{equation}
\begin{aligned}
\ev{\aac\aaa} &\to \Tr[\dop(t)\aac\aaa]/\Tr[\dop(0)] \\
  &= \ev{\aac\aaa} - i \eta \ev{\comm{\hat S(t)}{\aac\aaa}} \,,
\end{aligned}
\end{equation}
so that $\delta\bbr$ is no longer block anti-diagonal as in Eq.\
\eqref{eq:pnfam-even-dr-omega}.  Instead it has the form
\begin{equation}
\delta\bbr(t) \equiv \begin{pmatrix}
P_{\pi\nu}(t) & X_{\pi\nu}(t) \\
-X^*_{\pi\nu}(t) & -P^*_{\pi\nu}(t)
\end{pmatrix} \,,
\end{equation}
with $P(t)$ and $X(t)$ proportional to matrix elements of $\hat S(t)$:
\begin{subequations}
\begin{gather}
P_{\pi\nu}(t) \equiv i (f_\nu - f_\pi) S^{11}_{\pi\nu}(t), \\
X_{\pi\nu}(t) \equiv i (1 - f_\nu - f_\pi) S^{20}_{\pi\nu}(t).
\end{gather}
\end{subequations}
(The matrices $S^{11}$ and $S^{20}$ arise from the quasiparticle representation
of the one-body operator $\hat S(t)$; see, e.g., the Appendix of
Ref.\ \cite{Ring05}.) When the external field is sinusoidal, we have 
\begin{subequations}
\begin{align}
P_{\pi\nu}(t) &= P_{\pi\nu}(\omega) \emiwt + Q_{\pi\nu}^*(\omega) \epiwt, \\
X_{\pi\nu}(t) &= X_{\pi\nu}(\omega) \emiwt + Y_{\pi\nu}^*(\omega) \epiwt,
\end{align}
\end{subequations}
and, finally, the frequency-dependent perturbed density for an odd nucleus in the EFA is
\begin{equation}
\delta\bbr(\omega) = \begin{pmatrix}
P_{\pi\nu}(\omega) & X_{\pi\nu}(\omega) \\
-Y_{\pi\nu}(\omega) & -Q_{\pi\nu}(\omega)
\end{pmatrix}.
\end{equation}

The use of the EFA ensemble doubles the number of pnFAM equations, from two to
four:
\begin{widetext}
\begin{subequations} \label{eq:odd-pnfam-equations}
\begin{gather}
X_{\pi\nu}(\omega) [(E_\pi + E_\nu) - \omega]
  = -(1-f_\nu-f_\pi) [\delta H^{20}_{\pi\nu}(\omega) + F^{20}_{\pi\nu}], \label{eq:x-efa} \\
Y_{\pi\nu}(\omega) [(E_\pi + E_\nu) + \omega]
  = -(1-f_\nu-f_\pi) [\delta H^{02}_{\pi\nu}(\omega) + F^{02}_{\pi\nu}], \label{eq:y-efa} \\
P_{\pi\nu}(\omega) [(E_\pi - E_\nu) - \omega]
  = -(f_\nu-f_\pi) [\delta H^{11}_{\pi\nu}(\omega) + F^{11}_{\pi\nu}], \label{eq:p-efa} \\
Q_{\pi\nu}(\omega) [(E_\pi - E_\nu) + \omega] = -(f_\nu-f_\pi) [\delta
H^{\overline{11}}_{\pi\nu}(\omega) + F^{\overline{11}}_{\pi\nu}].
\label{eq:q-efa}
\end{gather}
\end{subequations}
\end{widetext}
Equations \eqref{eq:odd-pnfam-equations} are coupled through the dependence of
the Hamiltonian matrix $\delta\bbh(\omega)$ on the perturbed density
$\delta\bbr(\omega)$.  Besides the two `core' equations for $X$ and $Y$, which
are modified from Eqs.\ \eqref{eqs:pnfam-eqs}, the EFA pnFAM includes equations
for the matrices $P$ \eqref{eq:p-efa} and $Q$ \eqref{eq:q-efa}, which describe
transitions of the odd quasiparticle(s).

The EFA pnFAM equations equations are actually no more difficult to solve than
the usual ones.  Eqs.\ \eqref{eq:odd-pnfam-equations} contain the additional
matrices labeled $11$ and $\overline{11}$, but, because we solve for $\bbH$
iteratively in in the single-particle basis and then transform to the
quasiparticle basis \cite{Mustonen14}, we multiply by the Bogoliubov matrix
$\bbW$ in Eq.\ \eqref{eq:W-def} to obtain these additional matrices.  A few
additional iterations may be needed to solve the linear response equations, but
that does not significantly increase computation time.

% Strength function and interpretation
We compute the strength function in odd nuclei in the same way as in even ones,
but because $\delta\bbr(\omega)$ is not block anti-diagonal, the valence
nucleon(s) affects $S(F;\omega)$ explicitly through $P$ and $Q$, aw well as
implicitly through $X$ and $Y$:
\begin{equation} \label{eq:odd-sf}
\begin{aligned}
S(F;\omega) = \sum_{\pi\nu} & \Big[
  F^{20 *}_{\pi\nu} X_{\pi\nu}(\omega) + F^{02 *}_{\pi\nu} Y_{\pi\nu}(\omega) \\
&+
  F^{11 *}_{\pi\nu} P_{\pi\nu}(\omega) + F^{\overline{11} *}_{\pi\nu} Q_{\pi\nu}(\omega) \Big].
\end{aligned}
\end{equation}
Equation \eqref{eq:odd-sf} can be obtained directly from the EFA expectation
value $\Tr[\hat F^\dagger \dop(t)]/\Tr[\dop(0)]$ (cf.\ Refs.\
\cite{Nakatsukasa07,Avogadro11,Mustonen14}) by requiring that $\delta\dop(t)$
vary sinusoidally.  

With the EFA ensemble, $S(F;\omega)$ is simply the average transition strength
from the equally occupied odd-$A$ ground states $\aac_\Lambda\khfb$ and
$\aac_{\bar\Lambda}\khfb$:
\begin{equation} \label{eq:odd-sf-avg}
S(F;\omega) = \frac{1}{2} [ S_\Lambda(F;\omega) + S_{\bar\Lambda}(F;\omega) ] \,.
\end{equation}
(See Eq.\ \eqref{eq:efa-avg-ev-eo}.)  The two EFA states include the
polarization of the core due to the valence nucleon, at least partially
\cite{Perez-Martin08}.  
%The strength function obeys the Ikeda sum rule, since
%the commutator $\comm{\hat O^\dagger}{\hat O}$ of a one-body operator is itself
%a one-body operator; the sum rule should be satisfied separately for the
%auxiliary states and thus is satisfied for the full solution.

% Compare to Borzov and Marketin
In Fig.\ \ref{fig:efa-bgt}, we plot the total Gamow-Teller strength function for
the proton-odd nucleus $^{71}$Ga.  (In our EFA calculation, $^{71}$Ga has a
slight deformation $\beta_2=-0.007$; our methods for extracting lab-frame
transition strength from a deformed intrinsic nuclear ensemble are presented in
the appendix). The top panel compares the strength functions obtained with the
EFA-pnFAM and the even-even pnFAM, artificially constrained to obtain the
correct odd particle number as suggested in Ref.\ \cite{Marketin15}.  The two
calculations apply the same Skyrme energy-density functional (SV-min) without
proton-neutron isoscalar pairing, but begin with distinct HFB calculations.  The
EFA calculation clearly includes important one-quasiparticle transition strength
near $E_\mathrm{QRPA} = 1$ MeV that is not present in the other calculation.
The bottom panel compares the EFA-pnFAM strength, this time as a function of
excitation energy in the daughter nucleus (shifted downward in energy by $\epn
\simeq 749$ keV---see Eq.\ \eqref{eq:ex} and the discussion around
Eq.\ \eqref{eq:qbeta}) with a finite Fermi system calculation from
Ref.\ \cite{Borzov95}.  Although the two calculations do not yield identical
strength functions, they clearly mirror one another, and both include low-energy
one-quasiparticle strength. 

Finally, the odd-$A$ formalism of Ref.\ \cite{krumlinde84}, used by the authors
of Ref.\ \cite{moller03}, is an approximate version of ours.  We would recover
{similar expressions to those in Ref.\ \cite{krumlinde84} by substituting a
separable Gamow-Teller interaction for the Skyrme interaction and dropping terms
beyond leading order in $P_{\pi\nu}$ and $Q_{\pi\nu}$.

\begin{figure}
\includegraphics[scale=1]{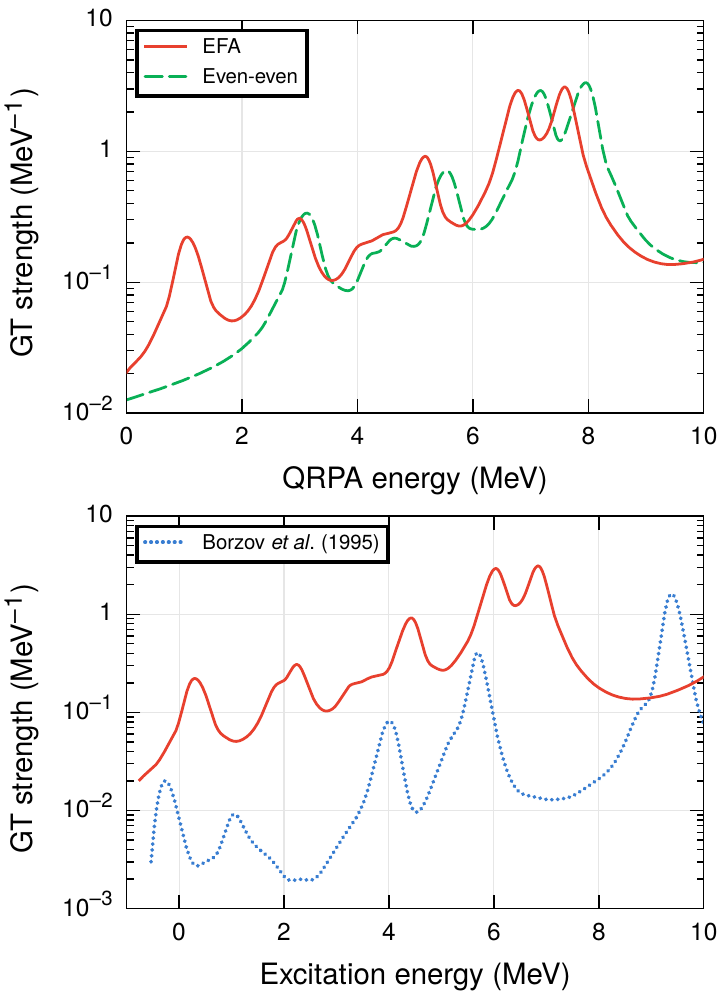}
\caption{\label{fig:efa-bgt}(Color online) \emph{Top panel:} Gamow-Teller
transition strength for $^{71}$Ga, computed with the EFA-pnFAM (red, solid
lines) and the even-even pnFAM (green, dashed lines), after constraining the
ground-state solution to have the correct odd average particle number, as in
Ref.\ \cite{Marketin15}.  \emph{Bottom panel:} The same EFA pnFAM strength
function as in the top panel, plotted vs.\ excitation energy $\ex =
E_\mathrm{QRPA}-\epn$ (see \eqref{eq:ex}), alongside the strength function from
Ref.\ \cite{Borzov95} (blue, dotted line).}
\end{figure}

% ------------------------------------------------------------------------------

\subsection{\label{sec:ns-beta}\boldmath Application to $\beta$ decay in deformed nuclei}

% We use ga=-1 everywhere
Ref.\ \cite{Mustonen14} discusses the calculation of $\beta$-decay rates from
pnFAM strength functions at length, so we make only a few important points here.
First, we treat the quenching of Gamow-Teller strength by using an effective
strength $g_A=-1.0$ for the axial-vector coupling constant, in both allowed and
first-forbidden $\beta$-decay transitions.  This renormalization is slightly
different from the that in Ref.\ \cite{Mustonen16}, where only Gamow-Teller
transitions were quenched.
% Q_beta
Second, we apply the $Q$-value approximation of Ref.\ \cite{engel99}:
\begin{equation} \label{eq:qbeta}
Q_\beta = \delmnh + \lambda_n - \lambda_p + \epn \,.
\end{equation}
Here $\delmnh$ is the neutron-Hydrogen mass difference, the $\lambda_q$ are
Fermi energies, and $\epn$ is the energy of the lowest two-quasiparticle state
for even-even nuclei, or the smallest one-quasiparticle transition energy for
odd nuclei.  We approximate $\epn$ with one-quasiparticle energies from the HFB
solution; this choice affects only $Q_\beta$, not the size of the QRPA energy
window, which is determined as in Ref.\ \cite{Mustonen14}:
\begin{equation}
E_\mathrm{QRPA}^\mathrm{max} = Q_\beta + \epn = \lambda_n - \lambda_p + \delmnh.
\end{equation}
Our procedure for going from the intrinsic frame to the lab frame, generalized
to include the odd-A EFA ensemble, is described in the appendix.  
% ------------------------------------------------------------------------------
% ------------------------------------------------------------------------------

\section{\label{sec:hl}Half-life calculations}

% ------------------------------------------------------------------------------

\subsection{\label{sec:hl-sensitivity}Identification of important nuclei}

To identify the most important $\beta$-decay rates for weak and rare-earth
\textit{r}-process nucleosynthesis, we turn to two sets of nucleosynthesis
sensitivity studies. Ref.\ \cite{mumpower16} reviews sensitivity studies for
main \textit{r} processes; our studies here proceed as described in Refs.\
\cite{mumpower2012c,Sur14,Mum14} and Section 5.2 of the review, Ref.\
\cite{mumpower16}.

The rare-earth peak is formed in a main \textit{r} process, so for our first set
of studies we begin with several choices of astrophysical conditions that
produce a good match to the solar \textit{r}-process pattern for $A\gtrsim120$.
These conditions include hot and cold parameterized winds, similar to those that
may occur in core-collapse supernovae or accretion disk outflows, along with
mildly heated neutron star merger ejecta.  We run a baseline simulation for each
astrophysical trajectory (i.e.\ condition) chosen, and then repeat it with
individual $\beta$-decay rates changed by a small factor, $K$.  Individual
$\beta$-decay half-lives in the rare-earth region tend to produce local changes
to the final abundance pattern that influence the size, shape, and location of
the rare-earth peak.  Thus we compare the final abundances with those of the
baseline simulation by using a local metric, $f_{local}$, defined as:
\begin{equation}\label{eqn:F}
f_{local}(Z,N)=100\times \sum_{A=150}^{180}|Y_{K}(A)-Y_{b}(A)| \,,
\end{equation}
where $Y_{b}$ is the final baseline isotopic abundance, and $Y_{K}$ is the final
abundance in a simulation in which the $\beta$-decay rate of the nucleus with
$Z$ protons and $N$ neutrons is multiplied by a factor $K$. Results for six
studies, in which individual $\beta$-decay rates were changed by a factor of
$K=5$, with hot, cold and neutron star merger \textit{r}-process conditions
appear in Fig.\ \ref{figure:beta_fgrid}.  The largest impacts to the final
abundances occur near the peak ($A\sim160$), though which nuclei are most
sensitive depends a little on the astrophysical conditions chosen.

\begin{figure}
\includegraphics[width=\linewidth]{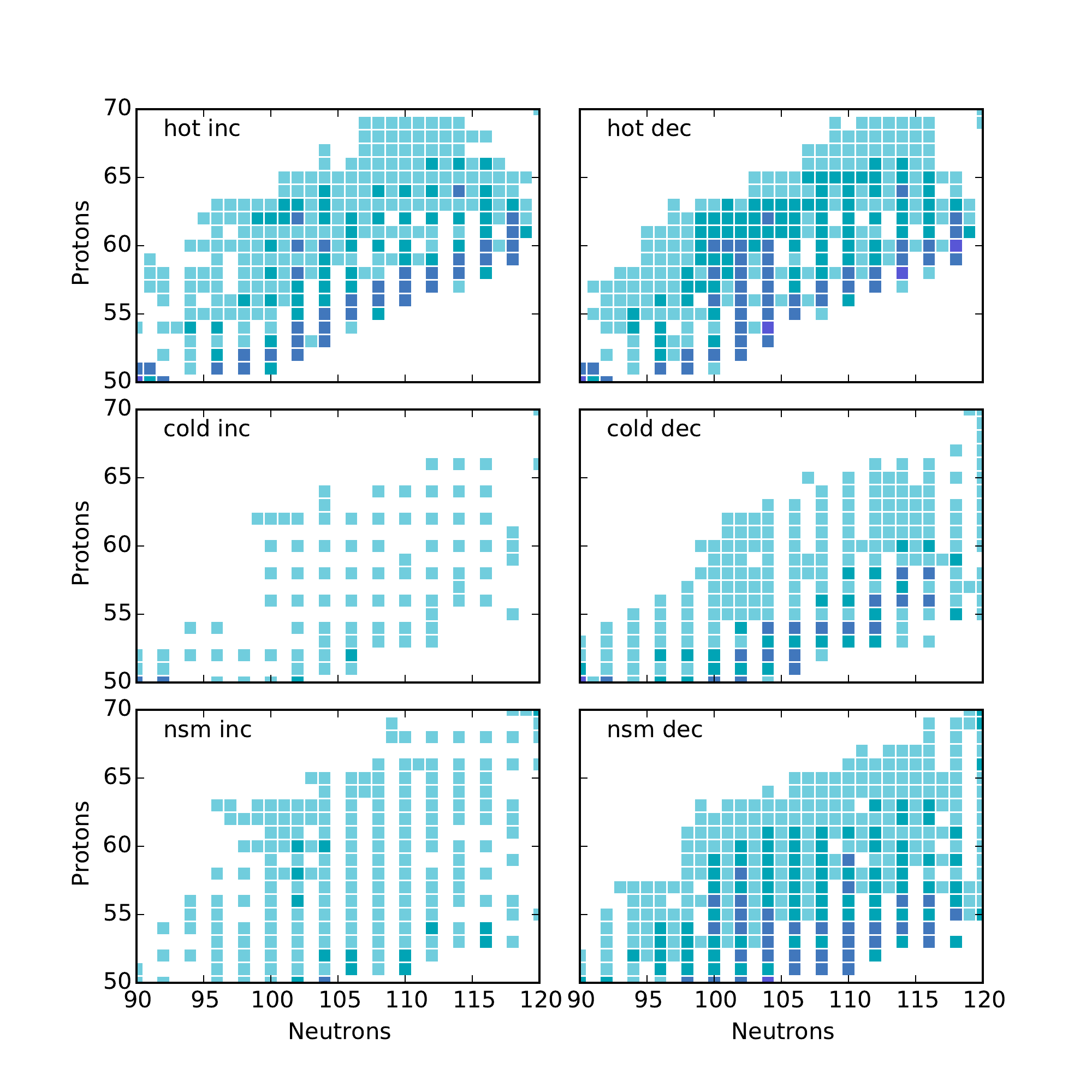}
\caption{\label{figure:beta_fgrid}(Color online) Influential $\beta$-decay rates
in the rare-earth region for hot, cold, and merger \textit{r}-process
conditions. The hot conditions are parameterized as in Ref.\ \cite{Mey02} with
entropy $s/k=200$, dynamical timescale $\tau_{\mathrm{dyn}}=80$ ms, and initial
electron fraction $Y_e=0.3$; the cold conditions are parameterized as in Ref.\
\cite{Panov2009} with $s/k=150$, $\tau_{dyn}=20$ ms, and $Y_e=0.3$; and the
merger conditions are from a simulation of A.\ Bauswain and H.-Th.\ Janka,
similar to that of Ref.\ \cite{Goriely2011}.  We performed two sensitivity
studies for each trajectory, looking at the results of increases and decreases
to the rates by a factor of $K=5$.  In order of lightest to darkest, the shades
are: white ($f_{local}=0$), light blue ($0.1<f_{local}\leq0.5$), medium blue
($0.5<f_{local}\leq1.0$), dark blue ($1<f_{local}\leq5$), and darkest blue
($f_{local}>5$).}
\end{figure}

In the second set of studies, focused on the $A\sim 80$ peak, we start with a
baseline weak \textit{r}-process simulation that produces an abundance pattern
with a good match to the solar pattern for $70<A<110$, as identified in Ref.\
\cite{Sur14}. Here we choose conditions qualitatively similar to those found in
the outflows of neutron star or neutron star-black hole accretion disks
\cite{Sur08,Wan14}, attractive candidate sites for the weak \textit{r} process.
We run a sensitivity study as described above, varying each $\beta$-decay
lifetime in turn by a factor of $K=10$ and comparing the result to the baseline
pattern. Unlike in the rare-earth region, where the influence of an individual
$\beta$-decay rate is primarily confined to the surrounding nuclei, rates in the
peak regions can produce global changes to the pattern \cite{Mum14} and can
influence how far the process proceeds in $A$.  Thus, in this study we compare
each pattern to the baseline with a global sensitivity measure $F_{global}$: 
\begin{equation}
F_{global}=100\times \sum_{A} |X_{K}(A)-X_{b}(A)| \,,
\label{eq:F}
\end{equation}
where $X_{b}(A)$ and $X_{K}(A)$ are the final mass fractions of the baseline
simulation and the simulation with the $\beta$-decay rate changed, respectively.
Figure \ref{fig:F} shows representative results. The pattern of most influential
$\beta$-decay lifetimes is similar to that identified for a main \textit{r}
process \cite{Mum14}: the important nuclei tend to be even-$N$ isotopes along
either the \textit{r}-process path or the decay pathways of the most abundant
nuclei.

\begin{figure}
\includegraphics[width=\linewidth]{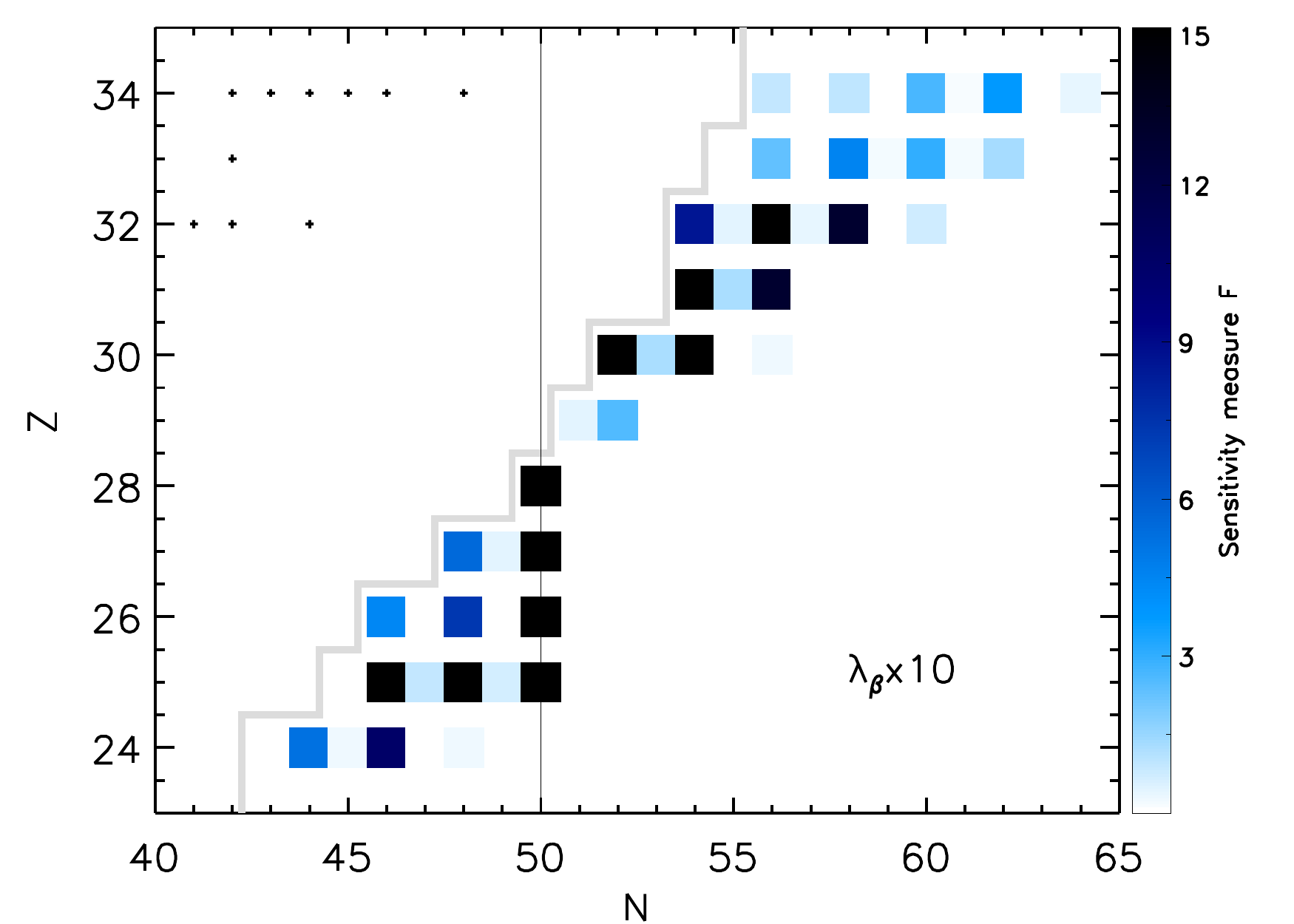}
\caption{\label{fig:F}(Color online) Influential $\beta$-decay rates in the
$A\sim 80$ region for weak \textit{r}-process conditions, parameterized as in
Ref.\ \cite{Mey02} with entropy per baryon $s/k=10$, dynamic timescale
$\tau=200$ ms, and starting electron fraction $Y_{e}=0.3$. The shaded boxes
show the global sensitivity measures $F_{global}$ resulting from $\beta$-decay
rate increases of a factor of $K=10$.  Stability is indicated by crosses; the
$\beta$-decay rates of nuclei to the right of stability and to the left of the
solid gray line have all been measured and so are not included in the
sensitivity analysis.}
\end{figure}

We select isotopic chains with the highest sensitivity measures, according to
Figs.\ \ref{figure:beta_fgrid} and \ref{fig:F}, and carefully recalculate their
$\beta$-decay half-lives.  The selected chains encompass 70 nuclei in the
rare-earth region and 45 nuclei in the $A\sim 80$ region \footnote{Measured
half-lives of $^{76,77}$Co and $^{80,81}$Cu were recently reported in
Ref.\ \cite{Xu14}. We still include these nuclei in our calculations.}.

% ------------------------------------------------------------------------------

\subsection{\label{sec:hl-edf}Selection and adjustment of Skyrme EDFs}

% Overview of the Skyrme EDF
Our density-dependent nucleon-nucleon interactions are derived from Skyrme EDFs.
Refs.\ \cite{Dobaczewski96b,Bender03,Perlinska04} contain comprehensive reviews
of the properties of Skyrme functionals; Refs.\ \cite{Mustonen14,Mustonen16}
contain discussions of the most important terms of the EDF for $\beta$ decay.
In our calculations, we largely apply the Skyrme EDF `as-is,' but adjust a few
important parameters that affect ground-state properties and $\beta$-decay
rates.  Among these are the proton and neutron like-particle pairing strengths,
$V_p$ and $V_n$; the spin-isospin coupling constant, $\cs$; and the
proton-neutron isoscalar pairing strength, $V_0$.  We tune these parameters
separately for each mass region; the coupling constants that multiply the
remaining `time-odd' terms of the Skyrme EDF are set either to values determined
by local gauge invariance \cite{Perlinska04} or to zero.

% ------------------------------------------------------------------------------

% Adjustment for the rare-earth nuclei
\subsubsection{\label{sec:hl-edf-ree}Multiple Skyrme EDFs for the rare-earth elements}

The pnFAM's efficiency significantly reduces the computational effort in
$\beta$-decay calculations.  The smaller computational cost makes repeated
calculations feasible and allows us to examine the extent to which $\beta$-decay
predictions depend on the choice of Skyrme EDF\@.  
% Selecting Skyrme EDFs and their properties
Here we use four very different Skyrme functionals: \skop \cite{Reinhard99},
SV-min \cite{Klupfel09}, \uneo \cite{UNE1}, and SLy5 \cite{Chabanat98}.  \skop,
has already been applied to the $\beta$ decay of spherical
nuclei \cite{engel99}; it was also chosen for the recent global calculations of
Ref.\ \cite{Mustonen16}.  SV-min, and \uneo are more recent; the latter is a
re-fit of the \textsc{unedf1} parameterization \cite{Kortelainen12}, without
Lipkin-Nogami pairing.  SLy5 tends to yield less-collective Gamow-Teller
strength than some other Skyrme parameterizations \cite{Fracasso07}.

We do most of our calculations in a 16-shell harmonic oscillator basis, a choice
that further reduces computational time from that associated with the 20-shell
basis applied in the \textsc{unedf} parameterizations of
Refs.\ \cite{Kortelainen10,Kortelainen12,Kortelainen13}.  Because \uneo was
constructed with \textsc{hfbtho} in a 20-shell basis, however, we use this
larger basis for that particular functional.  We determine the nuclear
deformation by starting from three trial shapes (spherical, prolate, and oblate)
and selecting the most bound result after the HFB energy and deformation have
been determined self-consistently.  We obtain the ground states of odd nuclei
within the EFA, beginning from a reference even-even solution and then computing
odd-$A$ solutions for a list of blocking candidates reported by \textsc{hfbtho}.
For odd-odd nuclei, we try all $N_p \times N_n$ proton-neutron configurations to
take into account as many odd-odd trial states as are practical.  Again, we
select the most-bound quasiparticle vacuum from among these candidates. 

% Overview of the fit details
Returning to the functionals themselves: to adjust the pairing strengths and
coupling constant $\cs$, we start from the published
parameterizations.\footnote{With a few exceptions: We use the same nucleon mass
for proton and neutrons, unlike Ref.\ \cite{Klupfel09}, which originally
determined SV-min, and we employ the SLy5 parameterization written into
\textsc{hfbtho}, which differs from that published in Ref.\ \cite{Chabanat98}.
The \textsc{hfbtho} values are the same as those of Ref.\ \cite{Bennaceur05},
but $t_0=-2483.45$ MeV~fm$^3$ instead of $-2488.345$ MeV~fm$^3$.}
%%%%% like-particle pairing
Then we fix the like-particle pairing strengths $V_p$ and $V_n$ by comparing the
average HFB pairing gap to the experimental odd-even staggering (OES) of nuclear
binding energies for the small set of test nuclei listed in Table
\ref{tab:re-isovector-dav}.\footnote{The \uneo pairing strengths were originally
fit simultaneously with the rest of the functional (with \textsc{hfbtho}), so we
do not re-adjust the \uneo pairing strengths.} Following the procedure in Refs.\
\cite{Kortelainen10,Kortelainen12,Kortelainen13}, we adjust the HFB pairing gap
to match the indicator (e.g., for neutrons) $\dav_n(Z,N) =
\tfrac{1}{2}[\ddd_n(Z,N+1) + \ddd_n(Z,N-1)]$ for even-even nuclei.  We obtain
the usual three-point indicators $\ddd$ \cite{Dobaczewski95,Satula98} from mass
excesses in the 2012 Atomic Mass Evaluation \cite{Audi12,Wang12}---after using
the prescription of Ref.\ \cite{Wang12} to remove the electron binding
\cite{Kortelainen10} from the atomic binding energies \cite{AME95}.  After
finding pairing strengths that correspond to one-$\sigma$ uncertainties in \dav
(treating asymmetric uncertainties as in Ref.\ \cite{Audi12b}), we find best-fit
values $V_p$ and $V_n$ for our sample set of nuclei.  For all EDFs except
SV-min, we choose mixed volume-surface pairing, with $\alpha=0.5$ as in Ref.\
\cite{Mustonen14}.  SV-min's pairing piece was originally fixed along with the
rest of the functional, but in the HF+BCS framework.  We therefore re-fit the
pairing strengths to better represent ground state properties with our HFB
solver, keeping the coefficient that specifies density dependence at its value
of $\alpha=0.75618$ from Ref.\ \cite{Klupfel09}.

\begin{table}
\caption{\label{tab:re-isovector-dav}%
OES indicators $\dav$ for the even-even nuclei used to fit the pairing
strengths $V_p$ and $V_n$. }
\begin{ruledtabular}
\begin{tabular}{r r D{,}{\,\pm\,}{-1} D{,}{\,\pm\,}{-1} }
$Z$ & $N$ & \multicolumn{1}{c}{$\dav_\mathrm{p}$ (MeV)} &
\multicolumn{1}{c}{$\dav_\mathrm{n}$ (MeV)} \\
\hline
52  &   84  &  0.79096,0.00464  &  0.75491,0.0024  \\
54  &   86  &  0.90975,0.01527  &  0.87276,0.0022  \\
56  &   90  &  0.92059,0.01093  &  0.92025,0.0204  \\
58  &   90  &  0.99503,0.00975  &  0.97777,0.0092  \\
60  &   92  &  0.68605,0.01176  &  0.77895,0.0298  \\
62  &   94  &  0.57543,0.02887  &  0.67368,0.0042  \\
62  &   96  &  0.55867,0.05021  &  0.58183,0.0049  \\
64  &   96  &  0.57608,0.00276  &  0.67969,0.0018  \\
66  &   98  &  0.53795,0.00277  &  0.67866,0.0016  \\
68  &  100  &  0.55392,0.00312  &  0.64734,0.0017  \\
68  &  102  &  0.50391,0.03646  &  0.60222,0.0021  \\
70  &  104  &  0.52725,0.00300  &  0.53483,0.0017  \\
72  &  106  &  0.62796,0.00168  &  0.63470,0.0016  \\
72  &  108  &  0.62486,0.00388  &  0.57799,0.0022  \\
74  &  110  &  0.55784,0.00199  &  0.66483,0.0008  \\
74  &  112  &  0.60795,0.01224  &  0.70165,0.0013  \\
74  &  114  &  0.67773,0.03607  &  0.79595,0.0227  \\
76  &  116  &  0.78248,0.01110  &  0.83218,0.0020  \\
78  &  118  &  0.75364,0.00128  &  0.88139,0.0009  \\
\end{tabular}
\end{ruledtabular}
\end{table}

%%%%% Cs, incl. SkO'-Nd
Next, we determine an appropriate value for $\cs$ by comparing the excitation energy,
\begin{equation} \label{eq:ex}
\ex = E_\mathrm{QRPA} - \epn \,,
\end{equation}
of the Gamow-Teller giant resonance (GTR) to an experimentally-measured value in
a nearby nucleus.  This constant $\cs$ is the same one we adjusted to GTR data
in the past \cite{Mustonen14}, following the work of Ref.\ \cite{Bender02};
Ref.\ \cite{Mustonen16} recently showed that it is the only particle-hole
constant that is truly important for $\beta$ decay.  In these \asimre nuclei, we
use the resonance associated with the doubly-magic nucleus $^{208}$Pb, with $\ex
= 15.6\pm 0.2$ MeV in the odd-odd daughter $^{208}$Bi \cite{Akimune95}, to fix
it.  We also use the deformed rare-earth nucleus $^{150}$Nd ($\ex \simeq 15.25$
MeV in $^{150}$Pm \cite{Guess11}), to fix an alternative value $\cs$ in \skop,
calling the resulting functional \ndsk.  The two fits result in values of $\cs$
that differ by nearly 20\%.  Figure \ref{fig:re-ndsk} compares the Gamow-Teller
strength functions produced by the two values in $^{150}$Pm.  Not only are the
resonances at different places, but there also a big difference in the strength
functions at the low energies that are important for $\beta$ decay. 

\begin{figure}
\includegraphics[scale=1]{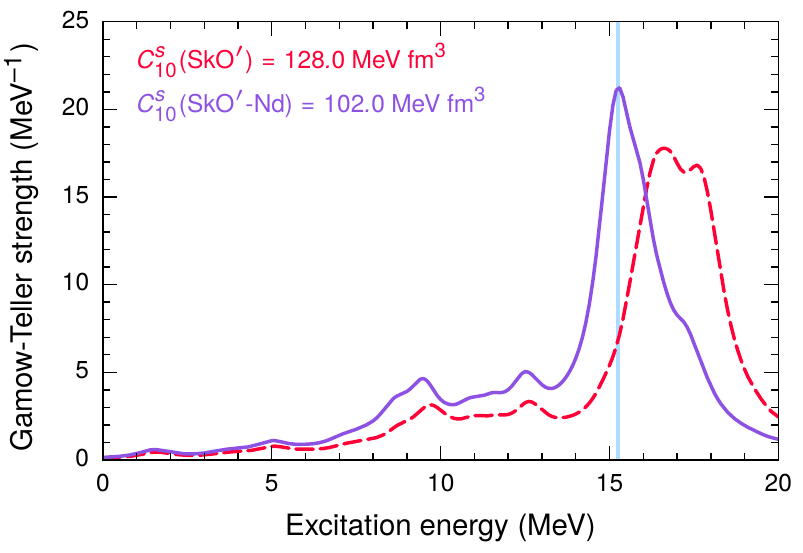}
\caption{\label{fig:re-ndsk}(Color online) Gamow-Teller strength functions in
$^{150}$Pm for \skop (red dashed line, with $\cs$ fit to the GTR energy in
$^{208}$Pb) and \ndsk (purple solid line, with $\cs$ fit to the GTR energy in
$^{150}$Nd).  The vertical line marks the measured GTR energy in $^{150}$Pm.}
\end{figure}

%%%%% V0, incl. difficulties
To adjust the $T=0$ pairing, we select short-lived even-even isotopes with $Z =
54$, 56, 58, 60, 62, and 64 with $\beta$-decay rates that have been measured
reasonably precisely, according to Ref.\ \cite{NWC081114-revtex}; the 18 nuclei
we use are listed in Table \ref{tab:re-isoscalar-fit}.  For each nucleus, we
attempt to find a pairing strength $V_0$ that reproduces the measured half-life.
If a calculated half-life is too short, even when $V_0=0$, we remove the nucleus
from consideration; this prevents our fit from being influenced by especially
long-lived or sensitive isotopes.  After determining an approximate $V_0 \ne 0$
for each nucleus (where possible), we compute average of these values, weighing
fast decays more than slow ones (since the very neutron-rich \textit{r}-process
nuclei are short-lived), with weight factors
\begin{equation} \label{eq:re-v0-weight}
w_i = \frac{1}{\log_{10} \big[ T_{1/2}^\mathrm{expt}(i) / 35\text{ ms} \big]}.
\end{equation}
The fit is fairly insensitive to the weighting half-life $T_0=35$ ms; with
$T_0=25$ ms the fit values of $V_0$ change by $\simeq2\%$.

% Analysis
Table \ref{tab:re-v0} lists the values for $V_0$ that we end up with and the
number of nuclei incorporated into the fit for each EDF\@.  We find that none of
the EDFs predict long-enough half-lives to fix $V_0 \ne 0$ for the entire set of
test nuclei; \skop (15 of 18) and SV-min (14) come the closest, while SLy5 (only
6) and \uneo (zero) come less close and are thus poorly constrained by $\beta$
decay.  (We discuss \ndsk momentarily.) One cannot really have confidence in
fits (SLy5, \uneo) that take into account less than half of the available data,
but Fig.\ \ref{fig:re-qvalues} provides at least a partial explanation.  It
compares our calculated $Q$ values \eqref{eq:qbeta} to measured values
\cite{Wang12} and those of the finite-range droplet model in Ref.\
\cite{moller97}.  Our $Q$ values are almost uniformly larger than experiment
(those of Ref. \cite{moller97} are generally smaller), and those of SLy5 and
\uneo are much larger.  Because the $\beta$-decay rate is roughly proportional
to $Q^5$ \cite{Basdevant05}, a $Q$ value that is too large will lead to an
artificially short half-life.  The $T=0$ pairing only make the half-lives
shorter.

\begin{table}
\caption{\label{tab:re-isoscalar-fit}%
Isotopes used to fit the proton-neutron isoscalar pairing to experimental
half-lives from Ref.\ \cite{NWC081114-revtex}.  Labels a--e in the ``Excluded?''
column note which isotopes were excluded from the fits for the functionals (a)
\skop, (b) \ndsk, (c) SV-min, (d) SLy5, and (e) \uneo, as discussed in the
text.}
\begin{ruledtabular}
\begin{tabular}{ r r r D{.}{.}{3} l}
$Z$ & $N$ & Element & \multicolumn{1}{c}{$T_{1/2}(\mathrm{expt})$} & Excluded? \\
\hline
54 &  88 & $^{142}$Xe  &   1.23   & \phantom{a} b \phantom{c} d e \\
54 &  90 & $^{144}$Xe  &   0.388  & \phantom{a} b \phantom{c} d e \\
54 &  92 & $^{146}$Xe  &   0.146  & \phantom{a} \phantom{b} \phantom{c} d e \\
56 &  88 & $^{144}$Ba  &  11.5    & a b \phantom{c} d e \\
56 &  90 & $^{146}$Ba  &   2.22   & \phantom{a} b \phantom{c} d e \\
56 &  92 & $^{148}$Ba  &   0.612  & \phantom{a} \phantom{b} \phantom{c} \phantom{d} e \\
58 &  90 & $^{148}$Ce  &   56     & \phantom{a} b c d e \\
58 &  92 & $^{150}$Ce  &    4     & \phantom{a} \phantom{b} \phantom{c} d e \\
58 &  94 & $^{152}$Ce  &    1.4   & \phantom{a} \phantom{b} \phantom{c} d e \\
60 &  92 & $^{152}$Nd  &  684     & a b c d e \\
60 &  94 & $^{154}$Nd  &   25.9   & \phantom{a} b c d e \\
60 &  96 & $^{156}$Nd  &    5.06  & \phantom{a} b \phantom{c} d e \\
62 &  96 & $^{158}$Sm  &  318     & a b c d e \\
62 &  98 & $^{160}$Sm  &    9.6   & \phantom{a} \phantom{b} \phantom{c} \phantom{d} e \\
62 & 100 & $^{162}$Sm  &    2.4   & \phantom{a} \phantom{b} \phantom{c} \phantom{d} e \\
64 &  98 & $^{162}$Gd  &  504     & \phantom{a} b \phantom{c} \phantom{d} e \\
64 & 100 & $^{164}$Gd  &   45     & \phantom{a} \phantom{b} \phantom{c} \phantom{d} e \\
64 & 102 & $^{166}$Gd  &    4.8   & \phantom{a} \phantom{b} \phantom{c} \phantom{d} e \\
\end{tabular}
\end{ruledtabular}
\end{table}

\begin{table}
\caption{\label{tab:re-v0}%
Summary of proton-neutron $T=0$ pairing fit, including the amount of test
data in each fit ($N$) and the resulting pairing strength ($V_0$).}
\begin{ruledtabular}
\begin{tabular}{l D{/}{/}{2} D{.}{.}{1}}
EDF  &  \multicolumn{1}{c}{$N$} &  \multicolumn{1}{c}{$V_0$} \\
\hline
\skop  & 15/18 & -320.0  \\
SV-min & 14/18 & -370.0  \\
\ndsk  &  9/18 & -300.0  \\
SLy5   &  6/18 & -240.0  \\
\uneo  &  0/18 &   -0.0  \\
\end{tabular}
\end{ruledtabular}
\end{table}

\begin{figure}
\includegraphics{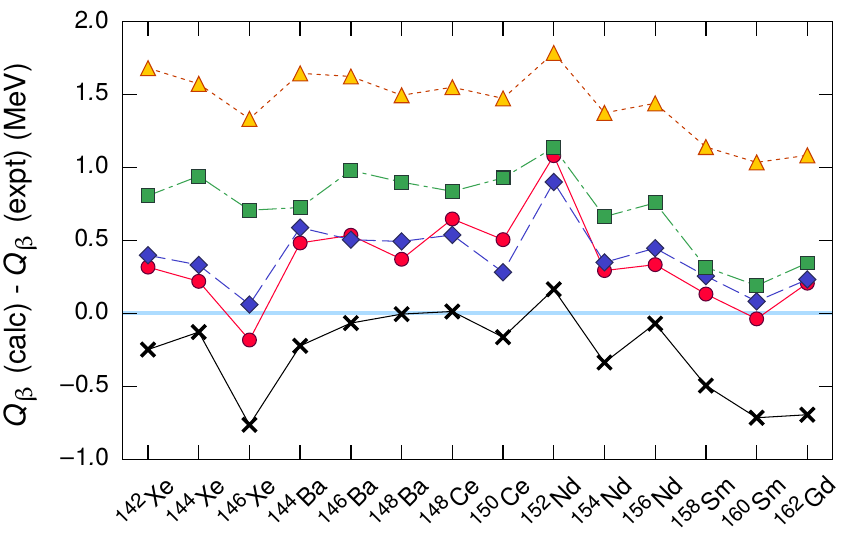}
\caption{\label{fig:re-qvalues}(Color online) The difference between calculated
and experimental $\beta$-decay $Q$ values, with \skop (circles), SV-min
(diamonds), SLy5 (squares), and \uneo (triangles).  $Q$ values of Ref.\
\cite{moller97} (crosses) also appear.}
\end{figure}

% Evaluation in even-even nuclei, w/discussion of EDFs and Q
The $Q$-value fitting difficulties, however, do not manifest themselves in
actual half-life predictions as much as they might, even with SLy5 and \uneo.
Figure \ref{fig:re-eval} compares our calculations to experimental measurements
in the 18 test nuclei used to fit $V_0$ (listed in Table
\ref{tab:re-isoscalar-fit}) and an additional 18 rare-earth nuclei (listed in
Table \ref{tab:re-isoscalar-test}).  Our results display the same pattern as
many others (e.g., Refs.\ \cite{moller03,Sarriguren15,Marketin15}), reproducing
half-lives of short-lived nuclei better than those of longer-lived ones.  The
$Q$-value errors discussed previously show up as systematic biases in our
half-life predictions (particularly with SLy5 and \uneo, for which half-lives of
long-lived nuclei are artificially reduced).  The shortest-lived nuclei,
however, are not so poorly represented even with SLy5 and \uneo; these
half-lives are still systematically short but by less than a factor of about two
(the shaded region in Fig.\ \ref{fig:re-eval} covers a factor of 5 relative to
measured values).  The systematic problems in the two functionals based on \skop
are barely noticeable.  SV-min, which performs the best overall, is somewhere in
the middle.  Because all the functionals do well with the short-lived isotopes,
we use them all for our rare-earth calculations.  The lifetimes we get with
\uneo serve as lower bounds on our predictions.

\begin{table}
\caption{\label{tab:re-isoscalar-test}%
The 18 even-even rare-earth nuclei in Fig.\ \ref{fig:re-eval} that are not
included in the EDF fitting.  Experimental
half-lives are from Ref.\ \cite{NWC081114-revtex}.}
\begin{ruledtabular}
\begin{tabular}{ r r r D{.}{.}{2}}
$Z$ & $N$ & Element & \multicolumn{1}{c}{$T_{1/2}(\mathrm{expt})$} \\
\hline
50 &  84 & $^{134}$Sn  &     1.05 \\
50 &  86 & $^{136}$Sn  &     0.25 \\
52 &  82 & $^{134}$Te  &  2508    \\
52 &  84 & $^{136}$Te  &    17.63 \\
52 &  86 & $^{138}$Te  &     1.4  \\
54 &  84 & $^{138}$Xe  &   844.8  \\
54 &  86 & $^{140}$Xe  &    13.6  \\
56 &  86 & $^{142}$Ba  &   636    \\
56 &  94 & $^{150}$Ba  &     0.3  \\
58 &  90 & $^{146}$Ce  &   811.2  \\
62 &  90 & $^{156}$Sm  &  33840   \\
66 & 102 & $^{168}$Dy  &    522   \\
68 & 106 & $^{174}$Er  &    192   \\
70 & 108 & $^{178}$Yb  &   4440   \\
70 & 110 & $^{180}$Yb  &    144   \\
72 & 112 & $^{184}$Hf  &  14832   \\
72 & 114 & $^{186}$Hf  &    156   \\
74 & 116 & $^{190}$W   &   1800   \\
\end{tabular}
\end{ruledtabular}
\end{table}

Finally, although \ndsk, the \skop variant that reproduces the GTR in
$^{150}$Nd, fails in 9 of the 18 nuclei used for fitting, its predictions do not
differ significantly from those of \skop.  Figure \ref{fig:re-ndsk} shows
increased low-energy Gamow-Teller transition strength as $\cs$ is reduced to
reproduce the rare-earth GTR\@.  This increased low-lying strength reduces
$\beta$-decay half-lives so that a smaller pairing strength is required (see
Table \ref{tab:re-v0}), with no loss of quality. 

\begin{figure}
\includegraphics{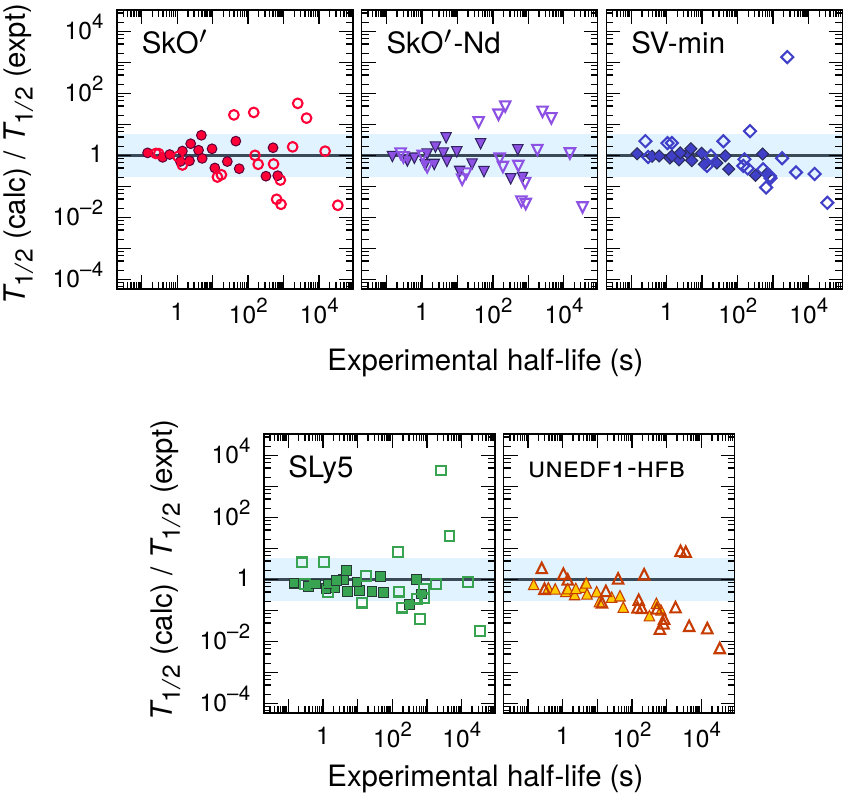}
\caption{\label{fig:re-eval}(Color online) Performance of fit functionals in
even-even rare-earth nuclei.  Filled symbols mark nuclei used to fit the $T=0$
pairing.  The experimental data are from Ref.\ \cite{NWC081114-revtex}.}
\end{figure}

% ------------------------------------------------------------------------------

% Adjustment for the A=80 nuclei
\subsubsection{\label{sec:hl-edf-a80}Weak \textit{r}-process elements}

% Choosing SV-min
To calculate the half-lives of \asimwk nuclei, we elect to apply the functional
from the previous section that best reproduces measured half-lives.  Both Fig.\
\ref{fig:re-eval} and the metrics in Ref.\ \cite{moller03} point to SV-min as
the best EDF\@.  We adjust SV-min for these lighter nuclei in much the same way
as discussed in the previous section.  We fit the like-particle pairing to
calculated OES values (see Table \ref{tab:a80-isovector-dav}), this time
employing the fitting software \pounders \cite{POUNDERS} to search for the best
values of $V_p$ and $V_n$. We provide \pounders with the weighted residuals
\begin{equation}
X_i = w_i \Big[
\bar\Delta_i(V_p,V_n) - \dav_i
\Big],
\end{equation}
where $\bar\Delta_i$ is the average HFB pairing gap for the $i$th nucleus and
the weight factor $w_i$ is 1 for non-magic nuclei and 10 for magic ones.  The
search yields $V_p=-361.0$ MeV fm$^3$ and $V_n=-320.9$ MeV fm$^3$. 
\begin{table}
\caption{\label{tab:a80-isovector-dav}%
\asimwk nuclei used to fit the like-particle pairing strengths.  We again use
data from Ref.\ \cite{Wang12} to compute the experimental indicators $\dav$. We
set $\dav=0$ for nuclei with $Z=28$ or $N=50$.}
\begin{ruledtabular}
\begin{tabular}{ c c D{.}{.}{7} D{.}{.}{7}}
$Z$ & $N$ & \multicolumn{1}{c}{$\dav_\mathrm{p}$ (MeV)} &
\multicolumn{1}{c}{$\dav_\mathrm{n}$ (MeV)} \\
\hline
   24  &  32  &  1.26354  &  1.01634 \\
   26  &  38  &  1.17387  &  1.29269 \\
   30  &  44  &  1.01199  &  1.41433 \\
   32  &  46  &  1.13235  &  1.24835 \\
   32  &  48  &  0.99635  &  1.17779 \\
   34  &  52  &  1.17100  &  0.78968 \\
   36  &  54  &  1.15773  &  0.83990 \\
   36  &  56  &  1.18243  &  0.90644 \\
   38  &  58  &  1.11437  &  0.93063 \\
   38  &  60  &  0.99089  &  0.85591 \\
   42  &  62  &  0.99786  &  0.95077 \\
   28  &  38  &  0.00000  &  1.20975 \\
   32  &  50  &  0.95788  &  0.00000 \\
   28  &  50  &  0.00000  &  0.00000 \\
\end{tabular}
\end{ruledtabular}
\end{table}
% C1s fit
We adjust the coupling constant $\cs$ to the GTR in a lighter doubly-magic
nucleus, $^{48}$Ca ($\ex = 10.6$ MeV in $^{48}$Sc \cite{Gaarde80}).

Finally, as before, we adjust the $T=0$ pairing strength to reproduce measured
half-lives of even-even nuclei, now with \asimwk (see Table
\ref{tab:a80-isoscalar-list}).  In this region of the isotopic chart the fit is
complicated by the presence of both proton and neutron closed shells (see Fig.\
\ref{fig:F}), so we include $Z=28$ and $N=50$ semi-magic nuclei in the fit.
Figure \ref{fig:a80-pn-fit} shows the impact of the $T=0$ pairing on half-lives
and in particular in the difference in the effect between non-magic nuclei
(solid lines) and semi-magic nuclei (dashed lines).  We search for distinct
values of $V_0$ for these two cases, finding $V_0(\mathrm{nm})=-353.0$
MeV~fm$^3$ for non-magic nuclei and $V_0(\mathrm{sm})=-549.0$ MeV~fm$^3$ is for
our set of semi-magic nuclei.

\begin{table}
\caption{\label{tab:a80-isoscalar-list}%
Open-shell (top) and semimagic (bottom) even-even \asimwk nuclei whose
half-lives are used to adjust the proton-neutron isoscalar pairing.
Experimental half-lives are from the ENSDF \cite{ENSDF}.}
\begin{ruledtabular}
\begin{tabular}{ r r r D{.}{.}{3}}
$Z$ & $N$ & Isotope & \multicolumn{1}{c}{$T_{1/2}$ (s)} \\
\hline
22  &  38  &  $^{60}$Ti  &  0.022 \\
24  &  40  &  $^{64}$Cr  &  0.043 \\
26  &  44  &  $^{70}$Fe  &  0.094 \\
30  &  52  &  $^{82}$Zn  &  0.228 \\
32  &  52  &  $^{84}$Ge  &  0.954 \\
34  &  54  &  $^{88}$Se  &  1.530 \\
36  &  60  &  $^{96}$Kr  &  0.080 \\
\hline
28  &  46  &  $^{74}$Ni  &  0.680 \\
28  &  48  &  $^{76}$Ni  &  0.238 \\
30  &  50  &  $^{80}$Zn  &  0.540 \\
32  &  50  &  $^{82}$Ge  &  4.560 \\
\end{tabular}
\end{ruledtabular}
\end{table}

\begin{figure}
\includegraphics{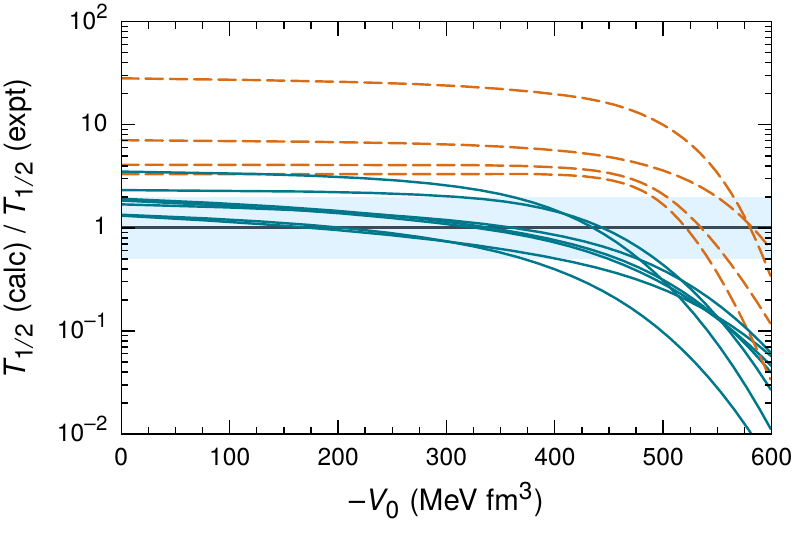}
\caption{\label{fig:a80-pn-fit}(Color online) Impact of $T=0$ pairing on
$\beta$-decay half-lives, both for non-magic (solid lines) and semi-magic
(dashed lines) nuclei.  The shaded region marks agreement between our
calculation and measured half-lives \cite{ENSDF} to within a factor of two.}
\end{figure}

The top panel of Fig.\ \ref{fig:a80-eval} shows the results of these
adjustments, comparing calculated $\beta$-decay half-lives of even-even (left
panel) and singly-odd (right) nuclei with measured values \cite{ENSDF}, for
nuclei with $22<Z<36$ and $T_{1/2}<1$ day.  The bottom panel shows the same
comparison for the finite-range droplet model (FRDM) calculation of Ref.\
\cite{moller03}.  The two sets of results are comparable, but those of Ref.\
\cite{moller03} have a clear bias in even-even nuclei.
\begin{figure}
\includegraphics{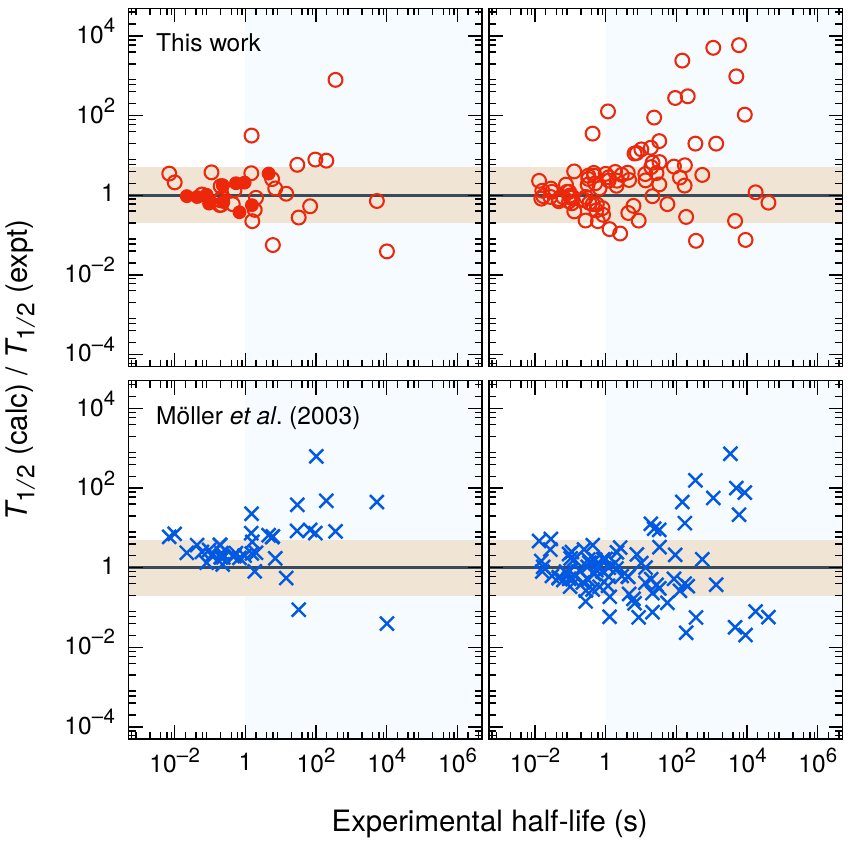}
\caption{\label{fig:a80-eval}(Color online) Performance of our SV-min
calculations (top panels) and the FRDM calculations of Ref.\ \cite{moller03}
(bottom panels) for nuclei with $22<Z<36$ and $T_{1/2}\le1$ day.  Left panels
show the ratio of calculated to measured half-lives in even-even nuclei; right
panels show the ratio in odd-$A$ nuclei.  Filled circles in the upper-left panel
denote even-even nuclei used to fit the $T=0$ proton-neutron pairing.  The
shaded horizontal band marks agreement to within a factor of five, and the white
background marks nuclei with measured half-lives that are shorter than one
second.}
\end{figure}
A metric defined in Ref.\ \cite{moller03}, 
\begin{equation}
r_i = \log_{10}\left[ \frac{T_{1/2}^\mathrm{calc}(i)}{T_{1/2}^\mathrm{expt}(i)} \right].
\end{equation}
captures the bias.  The mean and RMS deviation of the tenth power of $r_i$,
called $M_r^{10}$ and $\Sigma_r^{10}$, quantify the deviation between
calculation and experiment:} A value $M_r^{10}=2$ would signify that
calculations produce half-lives that are too long by a factor of two, on
average.  Our calculated rates yield $M_r^{10}=1.32$ in even-even nuclei while
those of Ref.\ \cite{moller03} give \ $M_r^{10}=3.55$.  For the standard
deviation, our rates yield $\Sigma_r^{10}=5.14$, vs.\ 7.50 for those of Ref.\
\cite{moller03}.  Thus, we indeed do measurably better in even-even nuclei.  Our
results in odd-$A$ nuclei are worse than those of Ref.\ \cite{moller03},
however; we obtain $M_r^{10}=2.71$ (vs.\ 0.95) and $\Sigma_r^{10}=11.61$ vs.\
(6.46).  The two sets of calculations are comparable for short-lived odd-$A$
nuclei, however: we get $M_r^{10}=1.11$ vs.\ 0.96 and $\Sigma_r^{10}=2.48$ vs.\
2.21 for isotopes with $T_{1/2} \le 1$ s.

% ------------------------------------------------------------------------------

\subsection{\label{sec:hl-ree}\boldmath Results near $A=160$}

\begin{figure}
\includegraphics{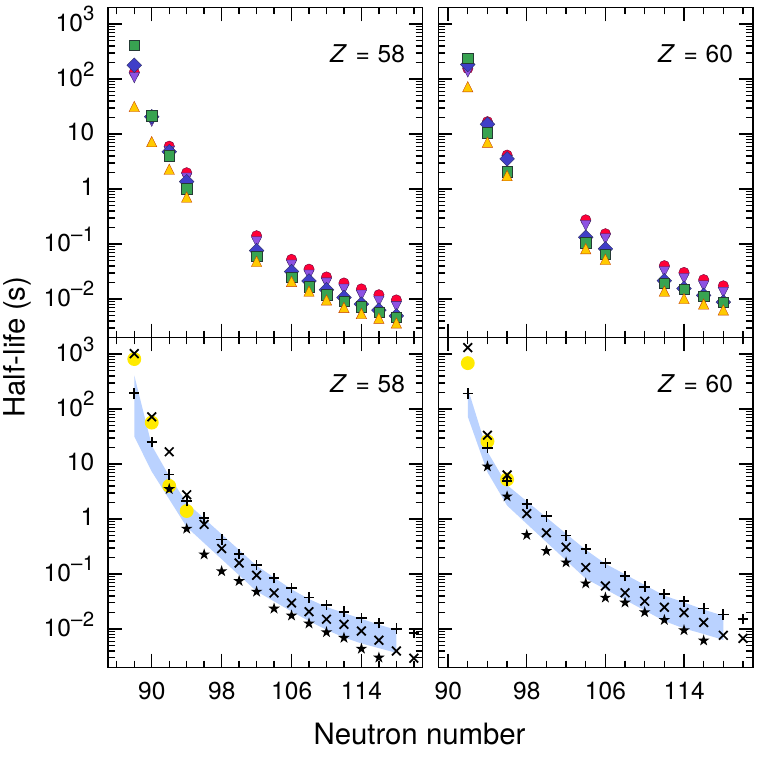}
\caption{\label{fig:re-hl}(Color online) \emph{Top panels:} Half-lives for
nuclei in the Ce ($Z=58$, left) and Nd ($Z=60$, right) isotopic chains,
calculated with the Skyrme EDFs described in the text.  Symbols correspond to
the same EDFs as in Fig.\ \ref{fig:re-eval}.  \emph{Bottom panels:} Calculated
half-lives of Refs.\ \cite{moller03} ($\times$), \cite{Mustonen16} ($+$), and
\cite{Fang16} ($\star$), measured half-lives \cite{ENSDF} (circles), and the
range of half-lives reported in this paper (shaded region).}
\end{figure}

% Results, w/similarity to related QRPA calculations
Guided by the sensitivity studies in Fig.\ \ref{figure:beta_fgrid}, we identify
70 rare-earth nuclei, all even-even or proton-odd, with rates that strongly
affect \textit{r}-process abundances near $A=160$.  (Neutron-odd nuclei do not
significantly affect the \textit{r} process since they quickly capture neutrons
to form even-$N$ isotopes \cite{BBFH}.) The top panels of Fig.\ \ref{fig:re-hl}
present new calculated half-lives in two isotopic chains, with all five adjusted
Skyrme EDFs.  The bottom panels compare our half-lives to measured values where
they are available \cite{NWC081114-revtex}, as well as to the results of
previous QRPA calculations \cite{moller03,Mustonen16,Fang16}.  Our calculations
span the (narrow) range of predicted half-lives in these isotopic chains, with
SLy5 and \uneo predicting the shortest half-lives for the most neutron-rich
isotopes, as one could expect from the analysis of Sec.\ \ref{sec:hl-edf-ree}.
While the \uneo half-lives are uniformly short, however, those of SLy5 actually
are actually the longest predictions (and the closest to measured values) for
nuclei nearer to stability.

%%%%% Discussion of Cs and Moller's results? Others?
The results of Refs.\ \cite{moller03,Mustonen16,Fang16} are actually fairly
similar to ours, spanning roughly the full range range of our predicted values.
The half-lives of Ref.\ \cite{Mustonen16} are close to our own \skop half-lives,
a result that is unsurprising given that the EDF in that paper is a modified
version of \skop (and that we use the same pnFAM code).  The half-lives of
Ref.\ \cite{moller03} lie, for the most part, right in the middle of our
predictions and follow those of SV-min fairly closely.  Finally, Fang's recent
calculations yield relatively short half-lives, shorter than even those of \uneo
most of the time.  Still, the band of predicted half-lives is relatively narrow
among these three calculations even in the most neutron-rich nuclei.
Ref.\ \cite{Mustonen16} points out that despite their differences, most global
QRPA calculations produce comparable half-lives. Our results in both \asimwk and
\asimre nuclei support this observation.

\begin{figure}
\includegraphics[scale=1]{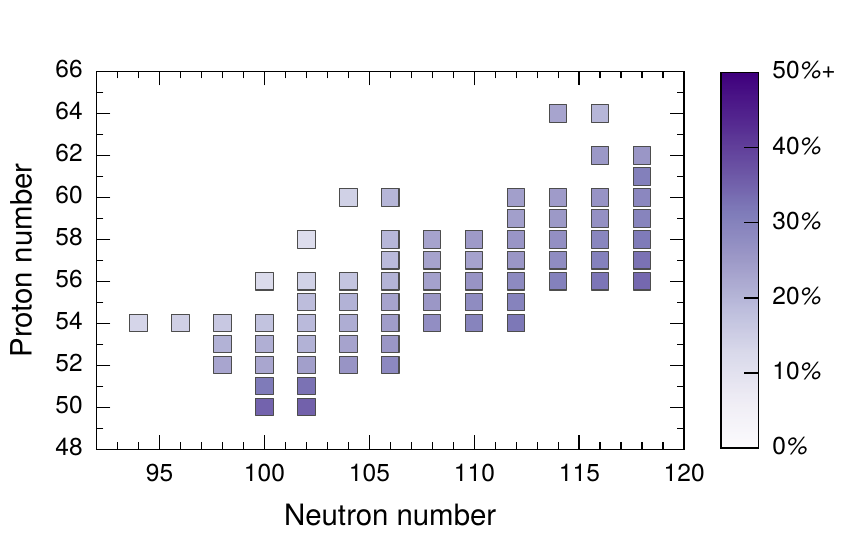}
\caption{\label{fig:re-ff-svmin}(Color online) Impact of first-forbidden
$\beta$ transitions in rare-earth nuclei that are important for the
\textit{r}-process nuclei, with the Skyrme EDF SV-min.}
\end{figure}

% FF contribution, cf. Marketin?
Finally, we have examined the impact of first-forbidden $\beta$ decay on
half-lives of rare-earth nuclei.  Figure \ref{fig:re-ff-svmin} shows that in
heavier nuclei forbidden decay makes up between 10 and 40 percent of the total
decay rate.  The percentage generally increases with $A$.

% ------------------------------------------------------------------------------

\subsection{\label{sec:hl-a80}\boldmath Results near $A=80$}

\begin{figure*}
\includegraphics{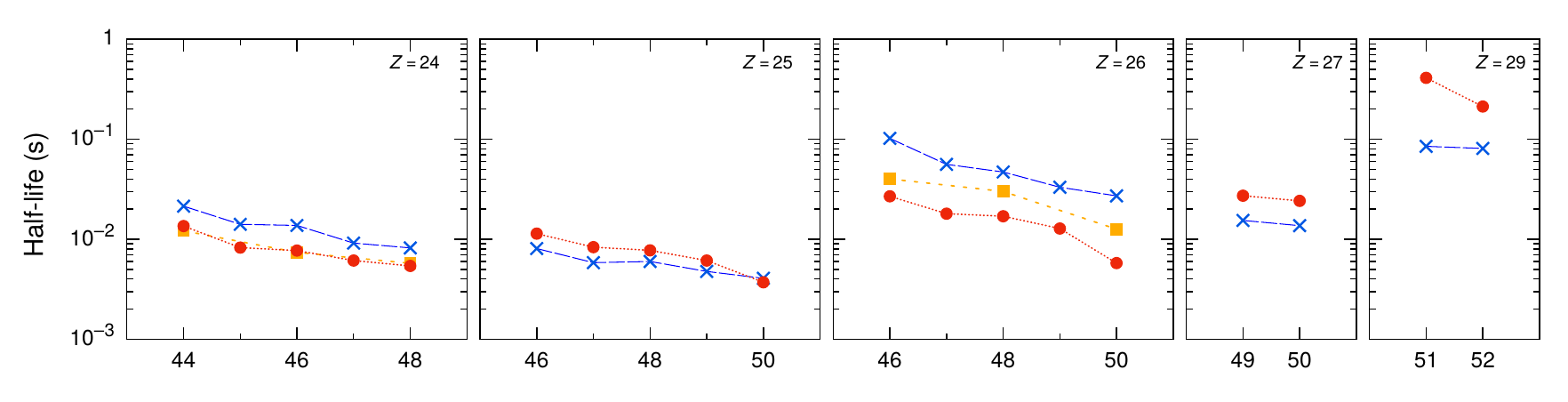}\\
\includegraphics{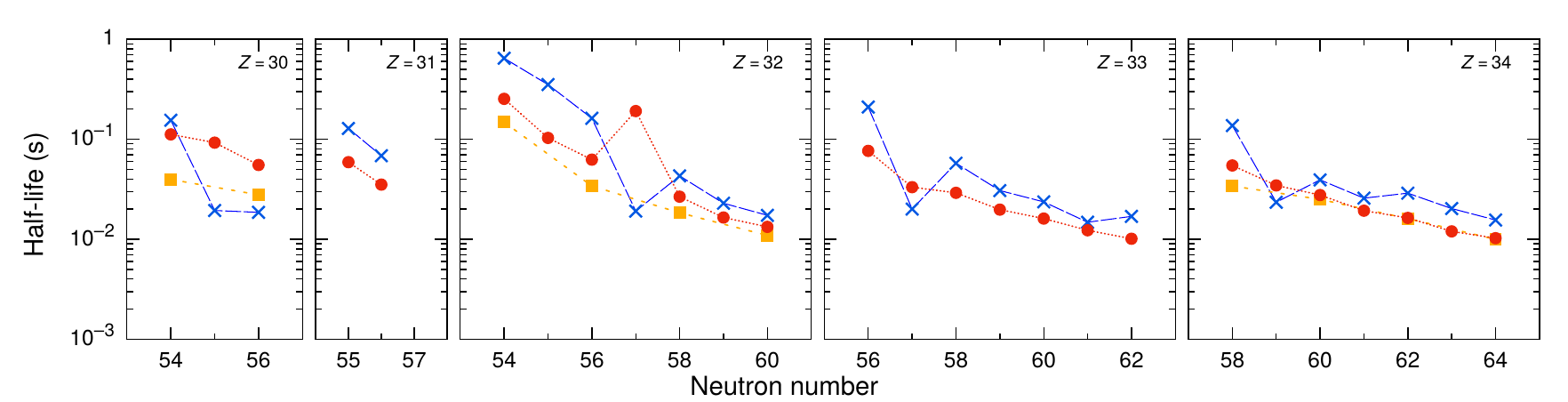}
\caption{\label{fig:a80-hl-final}(Color online) Computed half-lives in \asimwk
$\beta$-decay (red circles), along with those of Refs.\ \cite{Mustonen16}
(orange squares) and \cite{moller03} (blue crosses).}
\end{figure*}

% A=80 half-lives
Following the weak \textit{r}-process sensitivity study in Fig.\ \ref{fig:F}, we
present new half-lives for 45 \asimwk nuclei in Fig.\ \ref{fig:a80-hl-final},
comparing our results to those of Refs.\ \cite{moller03,Mustonen16}.  Not
surprisingly, in light of Fig.\ \ref{fig:a80-eval}, our calculated half-lives
(circles) are often slightly shorter than those of Ref.\ \cite{moller03}
(crosses).  They are also similar to those of Ref.\ \cite{Mustonen16}, which
used the same pnFAM code for even-even nuclei.  We have also compared our
\asimwk half-lives to those of the QRPA calculations in Ref.\
\cite{Sarriguren15}, finding very similar results for the few isotopic chains
discussed both here and there. 

\begin{figure}
\includegraphics{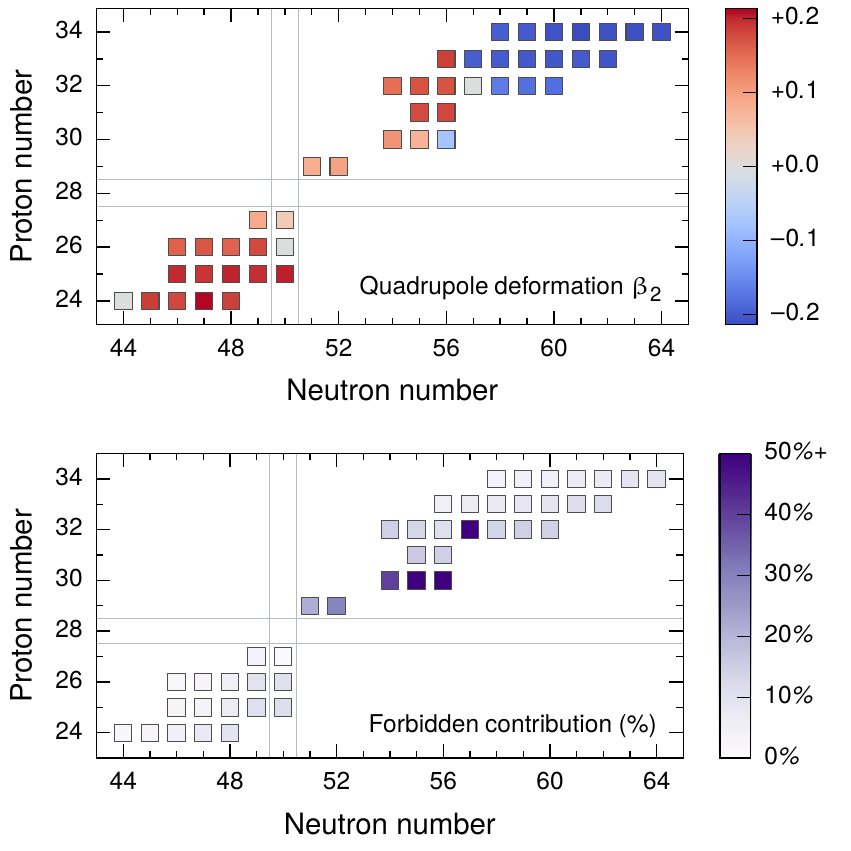}
\caption{\label{fig:a80-ff-contribs}(Color online) \emph{Top panel:} Quadrupole
deformation $\beta_2$ of the 45 \asimwk nuclei whose half-lives we calculate and
display in Fig.\ \ref{fig:a80-hl-final}.  \emph{Bottom panel:} Contribution of
first-forbidden $\beta$ decay ($0,1,2^-$ transitions) to the decay rate of the
same 45 nuclei, plotted as a percentage.}
\end{figure}

% Features---deformation and ff contribution
One interesting feature of our calculation is that the half-lives of
$^{85,86}$Zn, $^{89}$Ge, and (to a lesser extent) $^{90}$As are long compared to
those of Ref.\ \cite{moller03}.  The top panel of
Fig.\ \ref{fig:a80-ff-contribs}, which plots our calculated quadrupole
deformation $\beta_2$ for \asimwk nuclei, suggests these longer half-lives are
at least partially due to changes in ground-state deformation.  The Zn isotopes
switch from being slightly prolate to oblate near $^{86}$Zn ($N=56$), while Ge
and As isotopes do the same near $N=57$. Our calculations find $^{89}$Ge to be
spherical and situated between two isotopes with $\beta_2 \sim \pm 0.17$.  The
authors of Ref.\ \cite{moller03} appear to force $^{83-90}$Ge to be spherical. 

The bottom panel of Fig.\ \ref{fig:a80-ff-contribs} shows the impact of
first-forbidden $\beta$ decay in this mass region; together with the top panel
it connects negative-parity transitions with ground-state deformation.  As
discussed in Ref.\ \cite{Borzov03}, first-forbidden contributions to
$\beta$-decay rates are small, except in the oblate transitional nuclei
discussed above.  These results largely agree with those of the recent global
calculation in Ref.\ \cite{Mustonen16}. 

We also find that while the most deformed nuclei decay almost entirely via
allowed transitions, spherical isotopes show a large scatter in the contribution
of forbidden decay.  More than 80 percent of the $^{89}$Ge decay rate is driven
by first-forbidden transitions.  This analysis may bear on the large
first-forbidden contributions near $N=50$ and $Z=28$ reported by Ref.\
\cite{Marketin15}, which restricted nuclei to spherical shapes.  Ref.\
\cite{Sarriguren15}, which considers the effect of deformation on Gamow-Teller
strength functions, suggests that it is important near this mass region. 

Our calculations near $A=80$ include even-even, odd-even, even-odd, and odd-odd
nuclei.  Though odd pnFAM calculations are no more computationally difficult
than even ones, as discussed previously, a few odd half-lives are probably less
reliable than their even counterparts.  The reason is that when $E_\pi - E_\nu <
0$, we occasionally obtain QRPA energies that are negative.  Our technique for
calculating the decay rate then fails because of the form of the pnFAM strength
function \cite{Mustonen14}: for every state at $E=\hbar \omega$ that has
$\beta^-$ strength, the pnFAM generates a state with $\beta^+$ strength
(actually the negative of that strength) at $E=-\hbar \omega$.  As a result, if
there are states with negative energy and $\beta^-$ strength, then when we apply
the residue theorem to obtain integrated strength \cite{Hinohara13} we include
(the negative of) spurious $\beta^+$.  Because that strength is at low energy,
it is strongly weighted by $\beta$-decay phase space.  Of the 45 rates we
calculate near $A=80$, those for $^{80,81}$Cu, $^{86}$Ga, $^{94}$As, and
$^{97}$Se may be artificially small.  This problem does not seem to appear in
the rare-earth region. 

% ------------------------------------------------------------------------------
% ------------------------------------------------------------------------------

\section{\label{sec:rp}Consequences for the \textit{r} process}

\begin{figure}
\includegraphics[width=\linewidth]{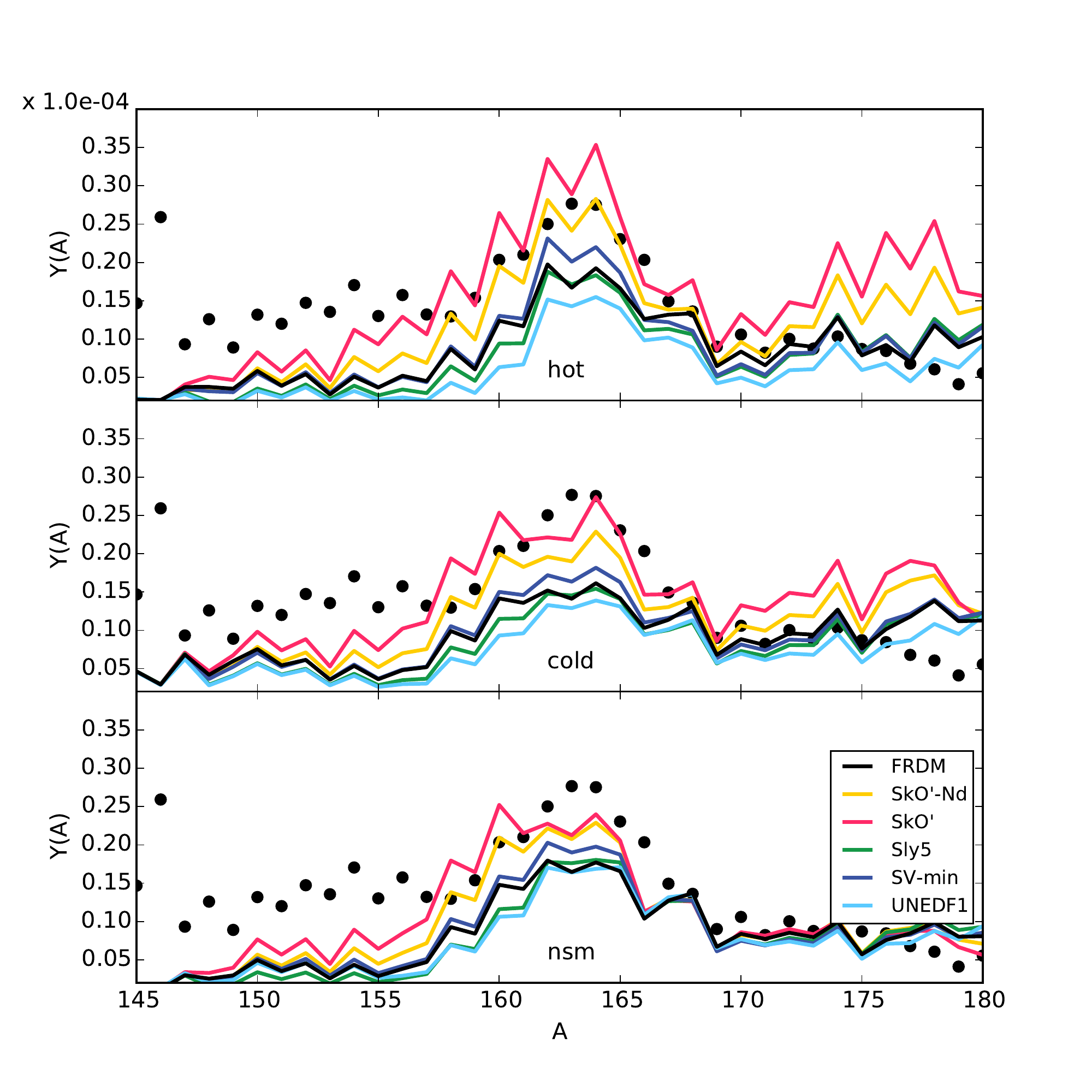}
\caption{\label{figure:ab_qrpa}(Color online) The effect of our new
$\beta$-decay rates on final \textit{r}-process abundances.  The same
trajectories are used as in Fig.\ \ref{figure:beta_fgrid}.}
\end{figure}

% Main r process
In rare-earth nuclei, our calculation produces rates that are either
consistently larger than or smaller than (depending on the functional) those of,
e.g., Ref.\ \cite{moller03}.  We now use these new rates in simulations of the
\textit{r} process.  The resulting rare-earth abundances are shown in Fig.\
\ref{figure:ab_qrpa}.  Generally, calculations that predict low rates build up
the rare-earth peak in both hot and cold \textit{r}-process trajectories, and
calculations that predict higher rates (SLy5 and \uneo) reduce the peak.  Our
neutron star merger calculation (bottom panel of Fig.\ \ref{figure:ab_qrpa})
demonstrates a different effect: longer-lived nuclei broaden the rare-earth
peak, and shorter-lived nuclei narrow it.  For all three trajectories the change
in abundances is fairly localized, with the effects caused by our most reliable
parameterizations (\skop, \ndsk, and SV-min) modifying abundances by factors,
roughly, of two to four.

\begin{figure}
\includegraphics[width=\linewidth]{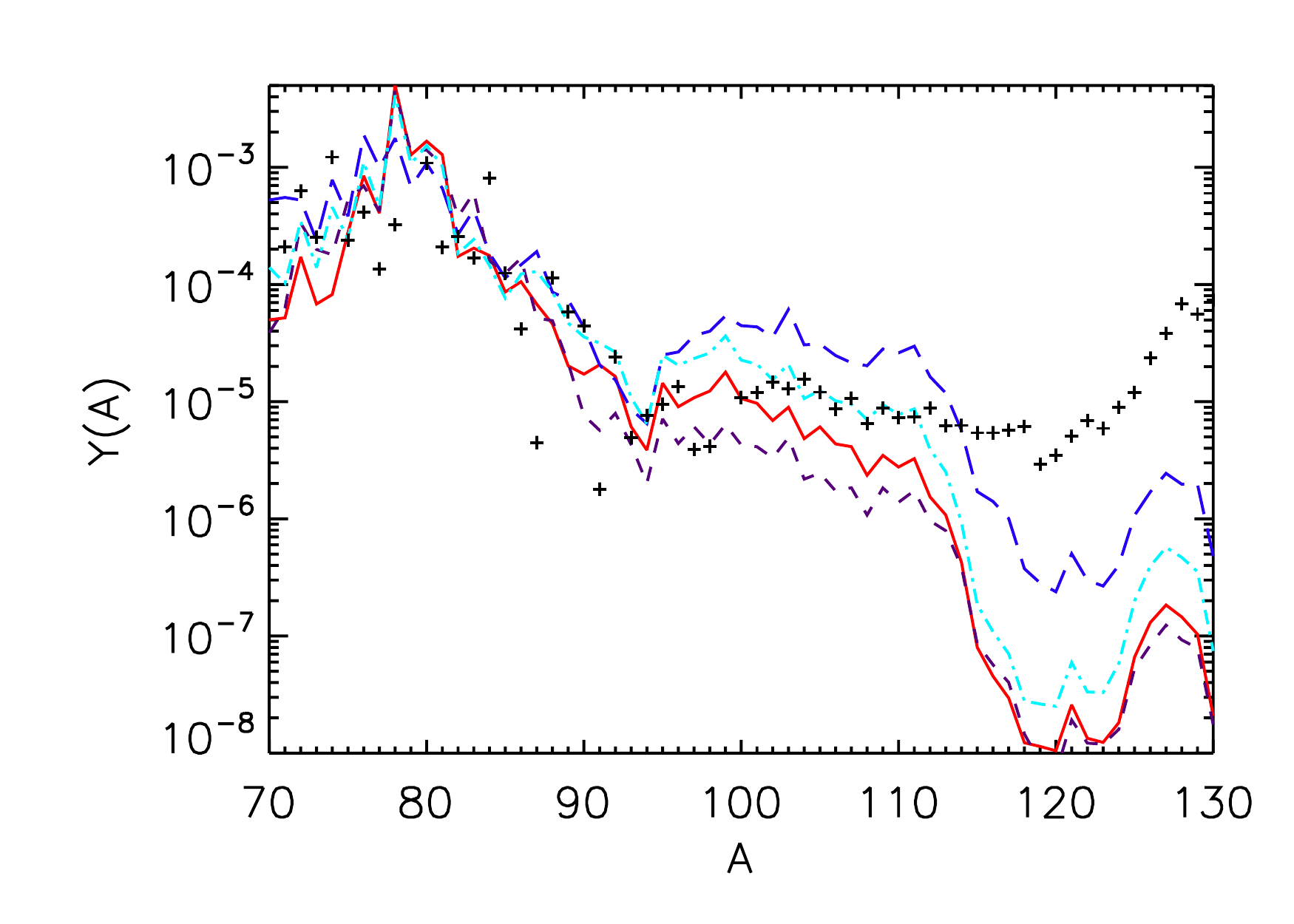}
\caption{\label{fig:a80-rp-effects}(Color online) Impact of our $\beta$-decay
rates near $A=80$ on weak \textit{r}-process abundances.  The top panel shows
abundances using rates from this work (red solid line), Ref.\ \cite{moller03}
(purple short dashes), Ref.\ \cite{Marketin15} (light blue dot dashes), and the
REACLIB database \cite{Cyburt10} (dark blue long dashes).}
\end{figure}

% Weak r process
Near $A=80$, our calculations produce small changes in weak \textit{r}-process
abundances from those obtained with the $\beta$-decay rates of Ref.\
\cite{moller03} and larger changes from those with compilations.  Fig.\
\ref{fig:a80-rp-effects} shows the baseline weak \textit{r}-process calculation
from the sensitivity study of Fig.\ \ref{fig:F} (blue line), where the
$\beta$-decay lifetimes are taken from the REACLIB database
\cite{Cyburt10}\footnote{\url{https://groups.nscl.msu.edu/jina/reaclib/db/}}
everywhere. We compare this abundance pattern to those produced when the
$\beta$-decay rates for the set of 45 nuclei calculated in this work are
replaced with our rates (red), those of Ref.\ \cite{moller03} (purple), and
those from Ref.\ \cite{Marketin15} (teal).  Although the differences in
abundance produced by our rates and those of Ref.\ \cite{moller03} are fairly
small, differences produced by ours and those of Ref.\ \cite{Marketin15} or
REACLIB are noticeable, with the widely-used REACLIB rates producing the most
divergent results.  It appears that many $\beta$-decay rates near $A=80$ in the
REACLIB database come from a much older QRPA calculation \cite{Klapdor84} that
differs significantly from the more modern calculations, especially in lighter
nuclei.

\begin{figure}
\includegraphics[width=3.25in]{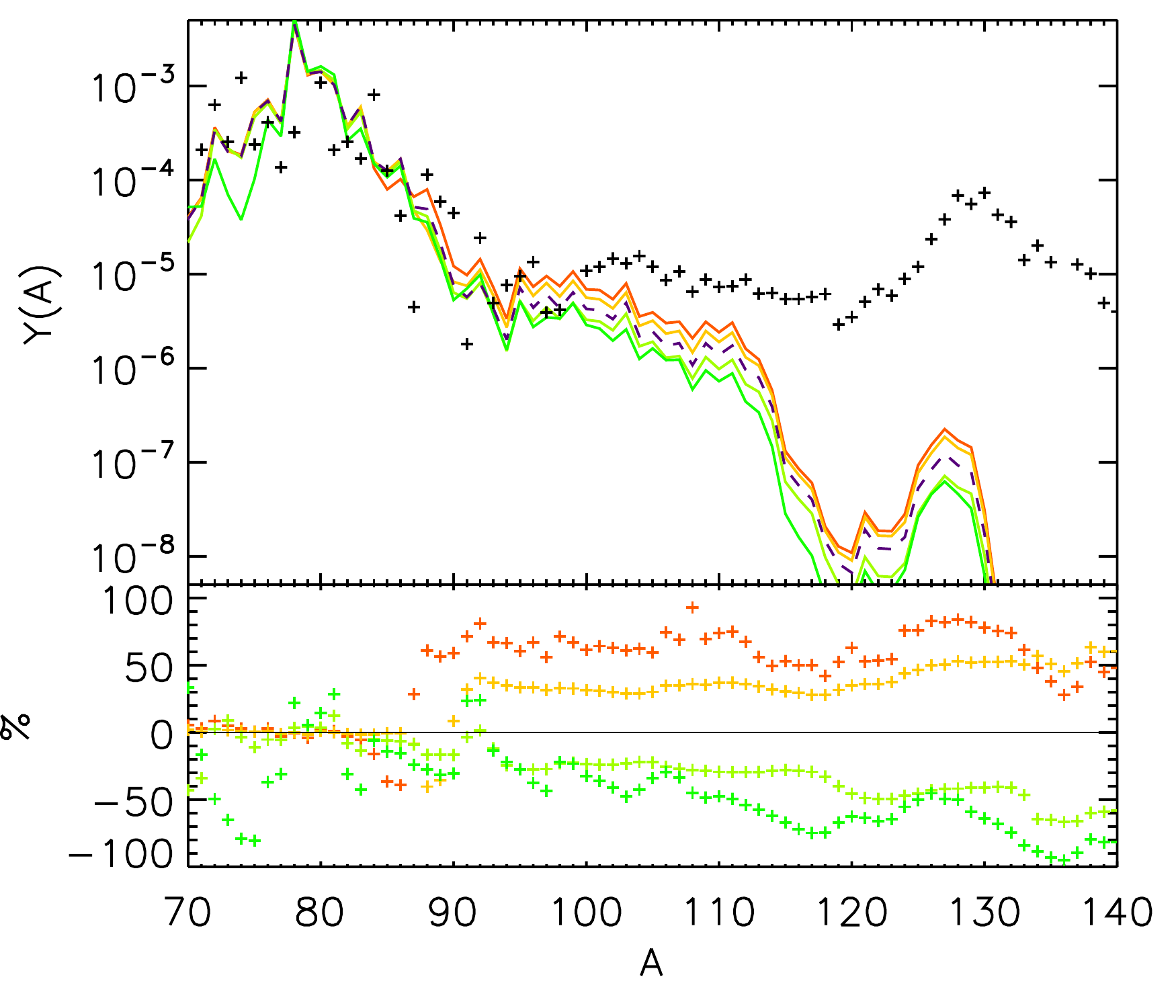}
\caption{\label{fig:yva}(Color online) \textit{Top panel:} Final abundance
pattern for the baseline trajectory described in the text, using the rates of
Ref.\ \cite{moller03} rates for all of the key nuclei identified in Sec.\ 2
(black line), compared to results of simulations with the same astrophysical
trajectory and with new rates for $^{68-72}$Cr only (light green line), new
rates for $^{72-76}$Fe only (green line), new rates for $^{86-92}$Ge only
(yellow line), and new rates for $^{89-95}$As only (orange line).
\textit{Bottom panel:} Percent difference between the abundances produced by the
baseline simulation (black line) and the simulations with the new rates (colored
lines).}
\end{figure}

Even though many of our calculated rates are higher than those of Ref.\
\cite{moller03} (Fig.\ \ref{fig:a80-hl-final}), their impact on a weak
\textit{r}-process abundance pattern is not a uniform speeding-up of the passage
of material through this region, as appears to be the case for for a main
\textit{r} process (see, e.g., Ref.\ \cite{engel99}).  $\beta$-decay rates can
influence how much neutron capture occurs in the $A\sim 80$ peak region and,
consequently, how many neutrons remain for capture elsewhere \cite{Mad12}.
Higher rates do not necessarily lead to a more robust weak \textit{r} process;
in fact, the opposite is more usually the case, since more capture in the peak
region generally leads to fewer neutrons available for capture above the peak.
This effect is illustrated in Fig.\ \ref{fig:yva}, which compares the abundance
pattern for the baseline weak \textit{r}-process simulation from Sec.\
\ref{sec:hl-sensitivity} with those obtained by using subsets of our newly
calculated rates in the same simulation.  Consider first the influence of the
rates of the iron isotopes (green line in Fig.\ \ref{fig:yva}), particularly
$^{76}$Fe.  This $N=50$ closed shell nucleus lies on the \textit{r}-process
path, below the $N=50$ nucleus closest to stability along the path, $^{78}$Ni.
Thus an increase to the $\beta$-decay rate of $^{76}$Fe over the its baseline
causes more material to move through the iron isotopic chain and reach the long
waiting point at $^{78}$Ni.  The abundances near the $A\sim 80$ peak increase
and those above the peak region decrease because the neutrons used to shift
material from the very abundant $^{76}$Fe to $^{78}$Ni are no longer available
for capture elsewhere.  Changes to the $\beta$-decay rates of nuclei just above
the $N=50$ closed shell, however, can have a quite different effect on the
pattern.  The germanium isotopes, particularly $^{86}$Ge and $^{88}$Ge, are just
above the $N=50$ closed shell, so increases to their $\beta$-decay rates from
the baseline will move material out of those isotopes to higher $A$ (orange line
in Fig.\ \ref{fig:yva}).  Thus, abundances above the peak increase and more
material makes it to the next closed shell, $N=82$.  In the end, the two very
different effects partly cancel one another so that our rates do not change
abundances significantly compared to those obtained with the rates of
Ref.\ \cite{moller03}.

\begin{figure}
\includegraphics[width=\linewidth]{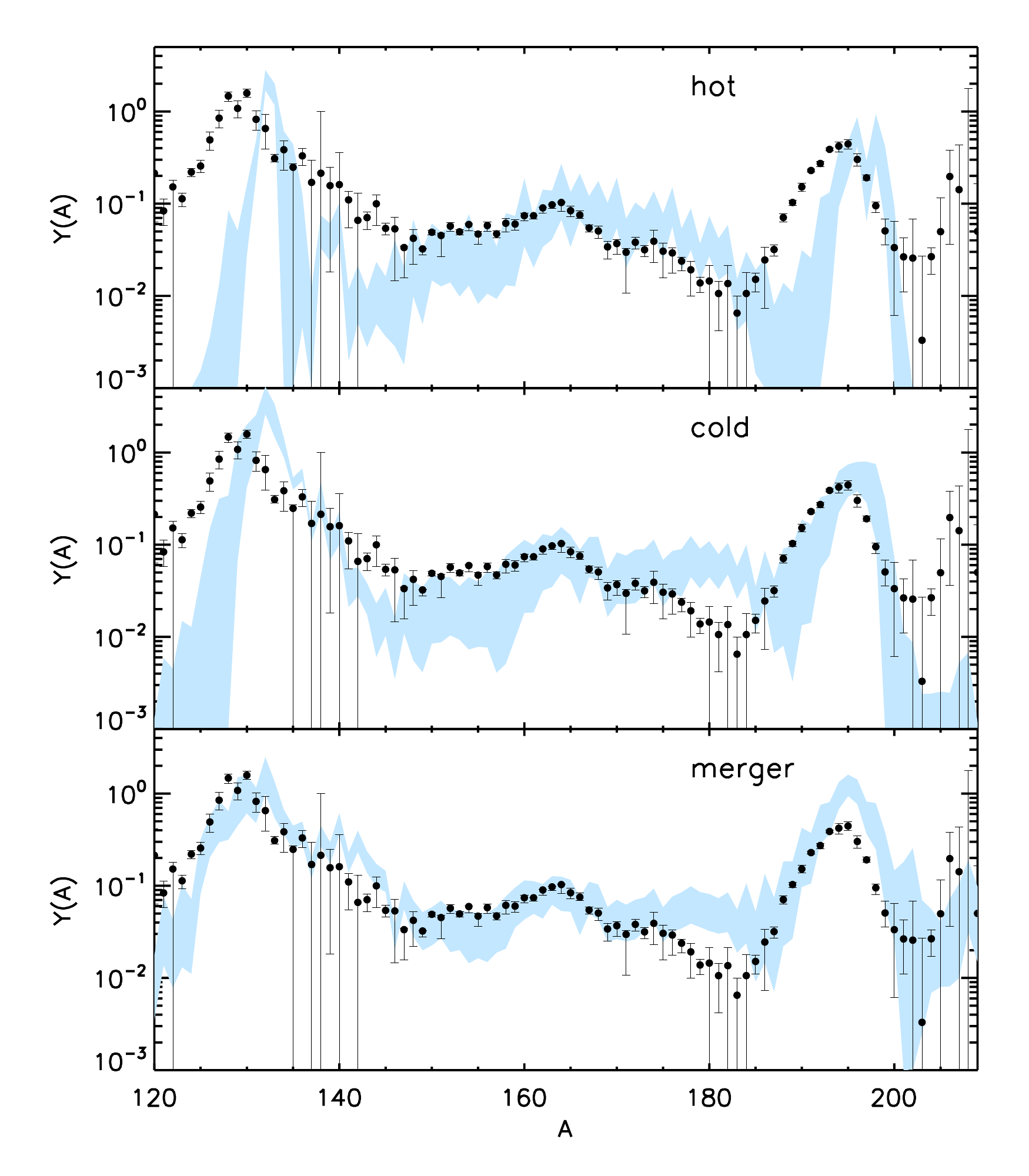}
\caption{\label{fig:MC}(Color online) Ranges of abundance patterns produced from
Monte Carlo sampling of all $\beta$-decay lifetimes from a log-normal
distribution of width two in the \textit{r}-process nuclear network, for hot
(top panel), cold (middle panel), and merger (bottom panel) trajectories as in
Ref.\ \cite{Surman+14}. Points are solar residuals from Ref.\ \cite{Arn07}.}
\end{figure}

Given that modern QRPA calculations appear to have converged to roughly a factor
of two or so, we use Monte Carlo variations to investigate the influence of this
amount of uncertainty in all the $\beta$-decay rates required for
\textit{r}-process simulations. We start with astrophysical trajectories that
seem typical for three types of main \textit{r}-process environments (hot wind,
cold wind, and merger). Then, for each Monte Carlo step, we vary all of the
$\beta$-decay rates by factors sampled from a log-normal distribution of width
two and re-run the \textit{r}-process simulation. Fig.\ \ref{fig:MC} shows the
resulting final \textit{r}-process abundance pattern variances for 10,000 such
steps. In each case, though some abundance pattern features stand out as clear
matches or mismatches to the solar pattern, the widths of the main peaks and the
size and shape of the rare-earth peak are not clearly defined.  The real
uncertainty in $\beta$-decay rates is larger than a factor of two because all
QRPA calculations miss what could be important low-lying correlations.  Thus,
more work is needed, whether it be theoretical refinement or advances in
experimental reach.

% ------------------------------------------------------------------------------
% ------------------------------------------------------------------------------

\section{\label{sec:conclusion}Conclusions}

We have adapted the proton-neutron finite-amplitude method (pnFAM) to calculate
the linear response of odd-$A$ and odd-odd nuclei, as well as the even-even
nuclei for which it was originally developed, by extending the method to the
equal-filling approximation (EFA).  The fast pnFAM can now be used to compute
strength functions and $\beta$ decay rates in all nuclei.  

After optimizing the nuclear interaction to best represent half-lives in each
mass region separately, we have calculated new half-lives for 70 rare-earth
nuclei and 45 nuclei near $A=80$.  Our calculated half-lives are broadly similar
to those obtained in the global calculations of \moetal.~\cite{moller03} as well
as to those of more recent work.  As a result, \textit{r}-process abundances
derived from our calculated half-lives are similar to those computed with the
standard rates of Ref.\ \cite{moller03}.  Our calculations support the
conclusions of Ref.\ \cite{Mustonen16}, which compared multiple QRPA
$\beta$-decay calculations and found that they all had similar predictions.
Still, the comparison of \textit{r}-process predictions with much older ones in
the REACLIB database and the discussion of uncertainty in Sec.\ \ref{sec:rp}
demonstrate the need for continued work on nuclear $\beta$-decay.

% ------------------------------------------------------------------------------
% ------------------------------------------------------------------------------

\begin{acknowledgments}

TS gratefully acknowledges many helpful conversations with M.~T.~Mustonen,
clarifying notes on SV-min from P.-G.~Reinhardt, a useful HFB fitting program
from N.~Schunck, and discussions with D.~L.~Fang.  This work was supported by
the U.S.\ Department of Energy through the Topical Collaboration in Nuclear
Science ``Neutrinos and Nucleosynthesis in Hot and Dense Matter,'' under
Contract No.~DE-SC0004142; through Early Career Award Grant No.~SC0010263 (CF);
and under individual contracts DE-FG02-97ER41019 (JE), DE-FG02-02ER41216 (GCM),
and DE-SC0013039 (RS). MM was supported by the National Science Foundation
through the Joint Institute for Nuclear Astrophysics grant numbers PHY0822648
and PHY1419765 and under the auspices of the National Nuclear Security
Administration of the U.S.\ Department of Energy at Los Alamos National
Laboratory under Contract No.  DE-AC52-06NA25396.  %
We carried out some of our calculations in the Extreme Science and Engineering
Discovery Environment (XSEDE)~\cite{XSEDE}, which is supported by National
Science Foundation grant number ACI-1053575, and with HPC resources provided by
the Texas Advanced Computing Center (\href{http://www.tacc.utexas.edu}{TACC}) at
The University of Texas at Austin.
\end{acknowledgments}

\appendix*
\section{Restoring Angular Momentum Symmetry}

% Deformation
% Basics
Deformed intrinsic states like those we generate in \textsc{hfbtho} require
angular-momentum projection.  Here we use the rotor model
\cite{Bohr98v1,Bohr75v2}, which is equivalent to projection in the limit of many
nucleons or rigid deformation \cite{Ring05}.  Even-even nuclei, with $K=0$
ground states ($K$ is the intrinsic $z$-component of the angular momentum) are
particularly simple rotors.  Their ``laboratory-frame'' reduced matrix elements
are just proportional to full intrinsic ones \cite{Bohr75v2},
\begin{equation} \label{eq:even-frame-trans}
\rmel{J\,K}{\hat O_J}{0\,0} = \Theta_K \mel{K}{\hat{O}_{JK}}{0}_\mathrm{intr},
\end{equation}
where $\Theta_K = 1$ for $K=0$ and $\Theta_K=\sqrt{2}$ for $K>0$.  The
corresponding transition strength to an excited state $\ket{J\,K}$ is 
\begin{equation}
B(\hat O_J;\, 0\,0\to J\,K) = \Theta_K^2 \abs{\mel{K}{\hat O_{JK}}{0}_\mathrm{intr}}^2.
\end{equation}
FAM strength functions are essentially composed of squared matrix elements
\cite{Nakatsukasa07,Hinohara13}, so we simply multiply $K>0$ strength functions
by $\Theta_{K}^2=2$.

% Odd nuclei: matrix element
The situation is more complicated in odd nuclei, which $K\neq 0$ ground-state
angular momenta.  The transformation between lab and intrinsic frames,
corresponding to Eq.\ \eqref{eq:even-frame-trans}, includes an additional term
involving the time-reversed intrinsic state
$\ket{\overline{K_i}}$ \cite{Bohr75v2}:
\begin{widetext}
\begin{equation} \label{eq:odd-frame-trans}
\begin{aligned}
\rmel{J_f\,K_f}{\hat O_{\lambda}}{J_i\,K_i} &= \sqrt{2J_i+1} \Big[
  \cg{J_i}{K_i;}{\lambda}{K_i-K_f}{J_f}{K_f}
  \mel{K_f}{\hat O_{\lambda,{K_f-K_i}}}{K_i}_\mathrm{intr} \\
& \qquad
+ \cg{J_i}{-K_i;}{\lambda}{K_i+K_f}{J_f}{K_f}
\mel{K_f}{\hat O_{\lambda,{K_f+K_i}}}{\overline{K_i}}_\mathrm{intr} \Big].
\end{aligned}
\end{equation}
\end{widetext}
% Odd nuclei: strength
If we neglect rotational energies in comparison with intrinsic energies, we can
sum over $J_i$ that appear in Eq.\ \eqref{eq:odd-frame-trans} to
obtain \cite{Bohr98v1}
\begin{widetext}
\begin{equation}\label{eq:b-odd}
\begin{aligned}
B(\hat O_\lambda;\, J_i\,K_i\to K_f) &=
  \frac{1}{2J_i+1} \sum_{J_f=\abs{J_i-\lambda}}^{J_i+\lambda}
  \abs{\rmel{J_f\,K_f}{\hat O_\lambda}{J_i\,K_i}}^2 \\
&= \abs{\mel{K_f}{\hat O_{\lambda,{K_i-K_f}}}{K_i}_\mathrm{intr}}^2
 + \abs{\mel{K_f}{\hat O_{\lambda,{K_i+K_f}}}{\overline{K_i}}_\mathrm{intr}}^2.
\end{aligned}
\end{equation}
\end{widetext}

% Example
The EFA-pnFAM, by preserving time-reversal symmetry and providing the combined
transition strength from auxiliary states $\aac_\Lambda\khfb$ and
$\aac_{\bar\Lambda}\khfb$, directly yields the terms in Eq.\ \eqref{eq:b-odd}.
For example, the Gamow-Teller decay of a nucleus with $J_i=K_i=3/2$ involves,
according to Eq.\ \eqref{eq:b-odd}, three intrinsic transitions:
\begin{equation*}
\mel{\tfrac 3 2}{\hat O_{K=0}}{\tfrac 3 2}, \quad
\mel{\tfrac 5 2}{\hat O_{K=1}}{\tfrac 3 2}, \quad
\mel{\tfrac 1 2}{\hat O_{K=-1}}{\tfrac 3 2}.
\end{equation*}
The third matrix element is equivalent because of time-reversal symmetry to
$\mel{{-\tfrac 1 2}}{\hat O_{K=1}}{{-\tfrac 3 2}}$.  Thus, a half-life
calculation to each band requires a $K=0$ transition and a pair of $K=1$
transitions from states $\ket{\tfrac 3 2}$ and $\ket{{-\tfrac 3 2}}$.  These are
the auxiliary states that make up the EFA-pnFAM strength function.  As a result,
we obtain the total strength to a band in an odd nucleus the same way as in an
even one: $S_\mathrm{total}(F;\omega) = S_{K=0}(F;\omega) + 2 S_{K=1}(F;\omega)$
for Gamow-Teller transitions.  The calculation of forbidden strength is similar.
The factor of two for $K>0$ strength cancels the factors of $1/2$ that appear in
the EFA strength function in \eqref{eq:odd-sf-avg}. $K=0$ transitions do not
need this factor since, e.g., $K=3/2 \to K=3/2$ and ${K=-3/2} \to {K=-3/2}$
transitions are equivalent and come together in the strength function.

% ------------------------------------------------------------------------------
% END BODY OF PAPER
% ------------------------------------------------------------------------------

%\bibliography{refs}
% \bibliography{rs,mm,trs1,trs2,trs3}  % RS, MM, dissertation intro/remainder, new refs

\end{document}